\documentclass[preprint,review,12pt,sort&compress]{elsarticle}%
\usepackage{amsmath}
\usepackage{geometry}
\usepackage{amssymb}
\usepackage{bbm}
\usepackage{amsthm}
\usepackage{amsfonts}
\usepackage{amsmath}
\usepackage{graphicx}
\usepackage[dvips,final]{epsfig}
\usepackage{setspace}
\usepackage{fancyhdr}
\usepackage{lineno}
\usepackage{caption}%
\providecommand{\U}[1]{\protect\rule{.1in}{.1in}}
\newcommand{\epspdo}[0]{\scriptsize\mathbf{\dot \varepsilon}_p = \mathbf{0}\normalsize\mathbf{}}
\newcommand{\epsdo}[0]{\scriptsize\mathbf{\dot \varepsilon} = \mathbf{0}\normalsize\mathbf{}}
\newcommand{\xpdo}[0]{\scriptsize\mathbf{\dot X}_p = \mathbf{0}\normalsize\mathbf{}}
\newcommand{\xdo}[0]{\scriptsize\mathbf{\dot X} = \mathbf{0}\normalsize\mathbf{}}

\newcommand{\ado}[0]{\scriptsize\mathbf{\dot A} = \mathbf{0}\normalsize\mathbf{}}
\newcommand{\lpo}[0]{\scriptsize\mathbf{l}_p = \mathbf{0}\normalsize\mathbf{}}
\newcommand{\lo}[0]{\scriptsize\mathbf{l} = \mathbf{0}\normalsize\mathbf{}}
\newcommand{\doo}[0]{\scriptsize\mathbf{d} = \mathbf{0}\normalsize\mathbf{}}
\newcommand{\edo}[0]{\scriptsize\mathbf{\dot E} = \mathbf{0}\normalsize\mathbf{}}

\newcommand{\lowpo}[0]{\scriptsize\mathbf{l} = \mathbf{0}; \mathbf{w}_p = \mathbf{0}\normalsize\mathbf{}}
\newcommand{\dowpo}[0]{\scriptsize\mathbf{d} = \mathbf{0}; \mathbf{w}_p = \mathbf{0}\normalsize\mathbf{}}
\renewcommand{\mathbf}[1]{\mbox{\boldmath $#1$ \unboldmath}  \!\!}
\journal{International Journal of Plasticity}
\begin{document}
\begin{frontmatter}
\title{A new class of plastic flow evolution equations for anisotropic
multiplicative elastoplasticity based on the notion of a corrector elastic strain rate}
\author[add1]{Marcos Latorre\corref{cor1}}
\ead{m.latorre.ferrus@upm.es}
\author[add1]{Francisco J. Mont\'{a}ns}
\ead{fco.montans@upm.es}
\address[add1]{Escuela T\'{e}cnica Superior de Ingenier\'{\i}a Aeron\'{a}utica y del Espacio\\Universidad Polit\'{e}cnica de Madrid\\
Plaza Cardenal Cisneros, 3, 28040-Madrid, Spain\\}
\cortext[cor1]{Corresponding author. Tel.:+34 913 366 367.}
\begin{abstract}
In this paper we present a new general framework for anisotropic elastoplasticity at large strains. The new framework
presents the following characteristics: (1) It is valid for non-moderate large strains, (2) it is valid for both elastic and
plastic anisotropy, (3) its description in rate form is parallel to that of the infinitesimal formulation, (4) it is compatible
with the multiplicative decomposition, (5) results in a similar framework in any stress-strain work-conjugate pair,
(6) it is consistent with the principle of maximum plastic dissipation and (7) does not impose any restriction on the
plastic spin, which must be given as an independent constitutive equation. Furthermore, when formulated
in terms of logarithmic strains in the intermediate configuration: (8) it may be easily integrated using a classical
backward-Euler rule resulting in an additive update. All these properties are obtained simply considering a plastic
evolution in terms of a corrector rate of the proper elastic strain. This formulation presents a natural framework for
elastoplasticity of both metals and soft materials and solves the so-called \emph{rate issue}.
\end{abstract}
\end{frontmatter}

\section{Introduction}

Constitutive models and integration algorithms for infinitesimal
elastoplasticity are well established
\cite{Simo98,KojicBatheBook,LublinerBook}. The currently favoured algorithmic
formulations, either Cutting Plane Algorithms or Closest Point Projection ones
are based on the concept of trial elastic predictor and subsequent plastic
correction \cite{SimoHughesBook}. The implementations of the most efficient
closest point projection algorithms perform both phases in just two subsequent
substeps \cite{MinCamMon}. From the 70's, quite a high number of formulations
have been proposed to extend both the continuum and the computational small
strain formulations to the finite deformation regime. Very different
ingredients have been employed in these formulations, as for example different
kinematic treatments of the constitutive equations, different forms of the
internal elastic-plastic kinematic decomposition, different types of stress
and strain measures being used, different internal variables chosen as the
basic ones and, most controversially, different evolution equations for the
plastic flow. The combinations of these ingredients have resulted into very
different extended formulations \cite{ShutovIhlemann14}. However, as a common
characteristic, all the formulations are developed with the main aim of
preserving as much as possible the simplicity of the classical return mapping
schemes of the infinitesimal theory
\cite{Wilkins64,MaenchenSacks64,KriegKey76} through an algorithm that computes
the closest point projection of the trial stresses onto the elastic domain.

The first strategies to model finite strain elastoplasticity were based on
both an additive decomposition of the deformation rate tensor into elastic and
plastic contributions and a hypoelastic relation for stresses
\cite{TruesdellNollBook}, see for example
\cite{HibbittEtAl70,McMeekingRice75,KeyKrieg82,TaylorBecker83} among many
others. Since the elastic stress relations are directly given in rate form and
do not derive in general from a stored energy potential, some well-known
problems may arise in these rate-form formulations, e.g. lack of objectivity
of the resulting integration algorithms and the appearance of nonphysical
energy dissipation in closed elastic cycles \cite{SimoPister84,KojicBathe87}.
Incrementally objective integration algorithms
\cite{HughesWinget80,PinskyOrtizPister83} overcome the former drawback; the
selection of the proper objective stress rate, i.e. the corotational
logarithmic rate in the so-called self-consistent Eulerian model
\cite{XiaoBruhnsMeyers97,BruhnsXiaoMeyers99,XiaoBruhnsMeyers00} circumvents
the latter one \cite{BrepolsVladimirovReese14}. Even though this approach is
still being followed by several authors \cite{Teeriaho13,XiaoEtAl16,ZhuEtAl16}
and may still be found in commercial finite element codes, the inherent
difficulty associated to the preservation of objectivity in incremental
algorithms makes these models less appealing from a computational standpoint
\cite{RubinsteinAtluri83,BrepolsVladimirovReese14}.

Shortly afterwards the intrinsic problems of hypoelastic rate models arose,
several hyperelastic frameworks formulated relative to different
configurations emerged \cite{ArgyrisDoltsinis79,SimoOrtiz85}. Green-elastic,
non-dissipative stresses are derived in these cases from a stored energy
function, hence elastic cycles become path-independent and yield no
dissipation \cite{GabrielBathe95}. Furthermore, objectivity requirements are
automatically satisfied by construction of the hyperelastic constitutive
relations \cite{SimoHughesBook}.

In hyperelastic-based models, the argument of the stored energy potential from
which the stresses locally derive is an internal elastic strain variable that
has to be previously defined from the total deformation. Two approaches are
common when large strains are considered. On the one hand, metric plasticity
models propose an additive split of a given Lagrangian strain tensor into
plastic and elastic contributions \cite{GreenNaghdi65}. On the other hand,
multiplicative plasticity models are based on the multiplicative decomposition
of the total deformation gradient into plastic and elastic parts \cite{Lee69}.
The main advantage of the former type is that the proposed split is parallel
to the infinitesimal one, where the additive decomposition of the total strain
into plastic and elastic counterparts is properly performed, so these models
somehow retain the desired simplicity of the small strain plasticity models
\cite{SimoOrtiz85,Miehe98,PapadopoulusLu98}. Another immediate consequence is
that these models are readily extended in order to include anisotropic
elasticity and/or plasticity effects
\cite{PapadopoulusLu01,Miehe02,LobleinSchroderGruttmann03,SansourWagner03,Ulz09}%
. However, it is well known that add hoc decompositions in terms of plastic
metrics do not represent correctly the elastic part of the deformation under
general, non-coaxial elastoplastic deformations
\cite{LublinerBook,Miehe02,GreenNaghdi71,Schmidt05}, hence its direct
inclusion in the stored energy function may be questioned. For example, it has
been found that these formulations do not yield a constant stress response
when a perfectly plastic isotropic material is subjected to simple shear, a
behavior which may be questionable \cite{Itskov04}. Furthermore, it has been
recently shown \cite{NeffGhiba16} that these formulations may even modify the
ellipticity properties of the stored energy function at some plastic
deformation levels, giving unstable elastic spring back computations as a
result, which seems an unrealistic response. On the contrary, multiplicative
plasticity models are micromechanically motivated from single crystal metal
plasticity \cite{Taylor38,Rice71}. The elastic part of the deformation
gradient accounts for the elastic lattice deformation and the corresponding
strain energy may be considered well defined. As a result, the mentioned
plastic shear and elastic spring back degenerate responses do not occur in
these physically sound models \cite{Itskov04,NeffGhiba16}.

Restricting now our attention to the widely accepted hyperelasto-plasticity
formulations based on the multiplicative decomposition of the deformation
gradient \cite{Lee69,Mandel74}, further kinematic and constitutive modelling
aspects have to be defined. On one side, even though spatial quadratic strain
measures were firstly employed \cite{Simo88}, they proved not to be natural in
order to preserve plastic incompressibility, which had to be explicitly
enforced in the update of the intermediate placement \cite{SimoMiehe92}. The
fact that logarithmic strain measures inherit some properties from the
infinitesimal ones, e.g. additiveness (only within principal directions),
material-spatial metric preservation, same deviatoric-volumetric projections,
etc., along with the excellent predictions that the logarithmic strain energy
with constant coefficients provided for moderate elastic stretches
\cite{Anand79,Anand86}, see also \cite{Rice75,RolphBathe84}, motivated the
consideration of the quadratic Hencky strain energy in isotropic
elastoplasticity formulations incorporating either isotropic or combined
isotropic-kinematic hardening
\cite{WeberAnand90,EterovicBathe90,PericOwenHonnor92,CuitinoOrtiz92,Simo92}.
Exact preservation of plastic volume for pressure insensitive yield criteria
is readily accomplished in this case. Moreover, the incremental schemes
written in terms of logarithmic strains preserve the desired structure of the
standard return mapping algorithms of classical plasticity models
\cite{Simo92}, hence providing the simplest computational framework suitable
for geometrically nonlinear finite element calculations.

On the other side, even though the use of logarithmic strain measures in
actual finite strain computational elastoplasticity models has achieved a
degree of common acceptance, a very controversial aspect of the theory still
remains. This issue is the specific form that the evolution equations for the
internal variables should adopt and how they must be further integrated
\cite{Bruhns15}, a topic coined as the \textquotedblleft rate
issue\textquotedblright\ by Sim\'{o} \cite{Simo92}. This issue originates,
indeed, the key differences between the existing models. In this respect, the
selection of the basic internal variable, whether elastic or plastic, in which
the evolution equation is written becomes fundamental in a large deformation
context. Evidently, this debate is irrelevant in the infinitesimal framework,
where both the strains and the strain rates are fully additive. Early works
\cite{Eckart48,Besseling66,Leonov76} suggest that the same strain variable on
which the material response depends, i.e. the internal elastic strains, should
govern the internal dissipation \cite{Rubin16}. This argument seems also
reasonable from a numerical viewpoint taking into account that in classical
integration algorithms \cite{Wilkins64,MaenchenSacks64,KriegKey76} the trial
stresses, which are elastic in nature and directly computed from the trial
elastic strains, govern the dissipative return onto the elastic domain during
the plastic correction substep. Following this approach, Sim\'{o}
\cite{Simo92} used a continuum evolution equation for associative plastic flow
explicitly expressed in terms of the Lie derivative of the elastic left
Cauchy--Green deformation tensor (taken as the basic internal deformation
variable \cite{SimoMiehe92}). He then derived an exponential return mapping
scheme to yield a Closest Point Projection algorithm formulated in elastic
logarithmic strain space identical in structure to the infinitesimal one,
hence solving the \textquotedblleft rate issue\textquotedblright%
\ \cite{Simo92}. However, the computational model is formulated in principal
directions and restricted to isotropy, so arguably that debated issue was only
partially solved. Extensions of this approach to anisotropy are scarce, often
involving important modifications regarding the standard return mapping
algorithms (cf. \cite{Rubin16} and references therein).

Instead, the probably most common approach when modeling large strain
multiplicative plasticity in the finite element context lies in the
integration of evolution equations for the plastic deformation gradient, as
done originally by Eterovic and Bathe \cite{EterovicBathe90} and Weber and
Anand \cite{WeberAnand90}. The integration is performed through an exponential
approximation to the incremental flow rule \cite{Simo98}, so these
formulations are restricted to moderately large elastic strains
\cite{EterovicBathe90,CamineroMontansBathe11}, which is certainly a minor
issue in metal plasticity. However we note that it may be relevant from a
computational standpoint if large steps are involved because the trial substep
may involve non-moderate large strains. Unlike Sim\'{o}'s approach, these
models retain a full tensorial formulation, so further consideration of
elastic and/or plastic anisotropy is amenable
\cite{ChattiEtAl01,HanEtAL03,EidelGruttman03,MenzelEtAl05,SansourKarsajSoric06,MontansBathe07,KimMontansBathe09,VladimirovPietrygaReese10,CamineroMontansBathe11}%
. However, the consideration of elastic anisotropy in these models has several
implications in both the continuum and the algorithmic formulations, all of
them derived from the fact that the resulting thermodynamical stress tensor in
the intermediate configuration, i.e. the Mandel stress tensor \cite{Mandel72},
is non-symmetric in general. Interestingly, the symmetric part of this stress
tensor is, in practice, work-conjugate of the elastic logarithmic strain
tensor for moderately large elastic deformations, which greatly simplifies the
algorithmic treatment \cite{CamineroMontansBathe11} in anisotropic metal
plasticity applications. As a result, the model in
\cite{CamineroMontansBathe11}, formulated in terms of generalized Kirchhoff
stresses instead of Kirchhoff stresses and with the additional assumption of
vanishing plastic spin, becomes the natural generalization of the Eterovic and
Bathe model \cite{EterovicBathe90} to the fully anisotropic case, retaining at
the same time the interesting features of the small strain elastoplasticity
theory and algorithms.

Summarizing, the computational model of Caminero et al.
\cite{CamineroMontansBathe11} is adequate for anisotropic elastoplasticity but
not for non-moderate large elastic deformations. In contrast, the Sim\'{o}
formulation \cite{Simo92} is valid for large elastic strains but not for
phenomenological anisotropic elastoplasticity. In this work we present a novel
continuum elastoplasticity framework in full space description valid for
anisotropic elastoplasticity and large elastic deformations consistent with
the Lee multiplicative decomposition. The main novelty is that, generalizing
Sim\'{o}'s approach \cite{Simo92}, internal elastic deformation variables are
taken as the basic variables that govern the local dissipation process. The
dissipation inequality is reinterpreted taking into account that the chosen
internal elastic tensorial variable depends on the respective internal plastic
variable and also on the external one. In this reinterpretation we take
special advantage of the concepts of partial differentiation and mapping
tensors \cite{LatMonAPM2016}. The procedure is general and may be described in
different configurations and in terms of different stress and strain measures,
yielding as a result dissipation inequalities that are fully equivalent to
each other. Respective thermodynamical \emph{symmetric} stress tensors and
associative flow rules expressed in terms of corrector elastic strain rates
and general yield functions are trivially obtained consistently with the
principle of maximum dissipation. We recover the Sim\'{o} framework from our
spatial formulation specialized to isotropy and with the additional assumption
of vanishing plastic spin, as implicitly assumed in Ref. \cite{Simo92}, see
also \cite{BonetWoodBook}. Exactly as it occurs in the infinitesimal theory,
in all the descriptions being addressed the plastic spin does not take
explicit part in the associative six-dimensional flow rules being derived,
hence bypassing the necessity of postulating a flow rule for the plastic spin
as an additional hypothesis in the dissipation equation \cite{Lubliner86}.
Special advantage is taken when the continuum formulation is written in terms
of the logarithmic elastic strain tensor \cite{LatMonIJSS2014}\ and its
work-conjugated generalized Kirchhoff \emph{symmetric} stress tensor, both
defined in the intermediate configuration. Then, the continuum formulation
mimics the additive description in rate form of the infinitesimal
elastoplasticity theory, the only differences coming from the additional
geometrical nonlinearities arising in a finite deformation context.
Furthermore, the \emph{unconventional appearance} \cite{Simo92} of the
well-known continuum evolution equation defining plastic flow in terms of the
Lie derivative of the elastic left Cauchy--Green tensor in the current
configuration \cite{BonetWoodBook} makes way for a\emph{ conventional}
evolution equation in terms of the elastic logarithmic strain rate tensor in
the intermediate placement, hence simplifying the continuum formulation to a
great extent and definitively solving the \textquotedblleft rate
issue\textquotedblright\ directly in the logarithmic strain space. Remarkably,
with the present multiplicative elastoplasticity model at hand, the generally
non-symmetric stress tensor that has traditionally governed the plastic
dissipation in the intermediate configuration, i.e. the Mandel stress tensor,
is no longer needed.

The rate formulation that we present herein in terms of logarithmic strains in
the intermediate configuration may be immediately recast in a remarkably
simple incremental form by direct backward-Euler integration which results in
integration algorithms of similar additive structure to those of the
infinitesimal framework. Indeed, the formulation derived herein is equivalent
in many aspects to the anisotropic finite strain viscoelasticity model based
on logarithmic strains and the Sidoroff multiplicative decomposition that we
presented in Ref. \cite{LatMonCM2015}. As done therein, a first order accurate
backward-Euler algorithm could be directly employed over the corrector
logarithmic elastic strain rate flow rule obtained herein to yield a return
mapping scheme in full tensorial form, valid for anisotropic finite strain
responses, that would preserve the appealing structure of the classical return
mapping schemes of infinitesimal plasticity without modification. For the
matter of simplicity in the exposition of the new elastoplasticity framework,
we do not include kinematic hardening effects in the formulation.
Nevertheless, its further consideration would be straightforward.

The rest of the paper is organized as follows. We next present in Section 2
the ideas for infinitesimal elastoplasticity in order to motivate and to
prepare the parallelism with the finite strain formulation. Thereafter we
present in Section 3 the large strain formulation in the spatial configuration
performing such parallelism. We then particularize the present proposal to
isotropy and demonstrate that some well-known formulations which are
restricted to isotropy are recovered as a particular case from the more
general, but at the same time simpler, anisotropic one. Section 4 is devoted
to the formulation in the intermediate configuration, where a comparison with
existing formulations is presented and some difficulties encountered in the
literature are discussed. Section 5 presents the new approach to the problem
at the intermediate configuration, both for quadratic strain measures and for
our favoured logarithmic ones. In that section we also discuss the advantages
and possibilities of the present framework.

\section{Infinitesimal elastoplasticity: two equivalent descriptions}

The purpose of this section is to motivate the concepts in the simpler
infinitesimal description, showing a new subtle view of these equations which,
thereafter result in a remarkable parallelism with the large strain formulations.

\begin{figure}
[ptb]
\begin{center}
\includegraphics[
height=1.2955in,
width=2.2442in
]%
{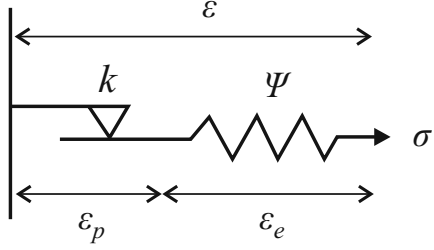}%
\caption{Rheological model motivating the (six-dimensional) elastoplasticity
model with (nonlinear) isotropic hardening.}%
\label{Figure - LeeSmall0.eps}%
\end{center}
\end{figure}

Consider the Prandtl (friction-spring) rheological model for small strains
shown in Figure \ref{Figure - LeeSmall0.eps}%
where $\mathbf{\varepsilon}$ and $\mathbf{\sigma}$ are the external,
measurable infinitesimal strains and engineering stresses, respectively, and
$\mathbf{\varepsilon}_{e}$ and $\mathbf{\varepsilon}_{p}$ are internal,
non-measurable infinitesimal strains describing the internal elastic and
plastic behaviors. The internal strains relate to the external ones through%
\begin{equation}
\mathbf{\varepsilon}=\mathbf{\varepsilon}_{e}+\mathbf{\varepsilon}_{p}
\label{eps=epse+epsp}%
\end{equation}
so if we know the total deformation and one internal variable, then the other
internal variable is uniquely determined. We will consider
$\mathbf{\varepsilon}$ and $\mathbf{\varepsilon}_{p}$ as the
\emph{independent} variables of the dissipative system and
$\mathbf{\varepsilon}_{e}$ will be the \emph{dependent} internal variable. The
following two-variable dependence emerges for $\mathbf{\varepsilon}_{e}$%
\begin{equation}
\mathbf{\varepsilon}_{e}\left(  \mathbf{\varepsilon},\mathbf{\varepsilon}%
_{p}\right)  =\mathbf{\varepsilon}-\mathbf{\varepsilon}_{p}
\label{epse dependence 0}%
\end{equation}
which provides also a relation between the corresponding strain rate
tensors---we use the notation $\partial\left(  \cdot\right)  /\partial(\circ)$
for partial differentiation%
\begin{equation}
\mathbf{\dot{\varepsilon}}_{e}=\left.  \frac{\partial\mathbf{\varepsilon}_{e}%
}{\partial\mathbf{\varepsilon}}\right\vert _{%
\epspdo
}:\mathbf{\dot{\varepsilon}}+\left.  \frac{\partial\mathbf{\varepsilon}_{e}%
}{\partial\mathbf{\varepsilon}_{p}}\right\vert _{%
\epsdo
}:\mathbf{\dot{\varepsilon}}_{p}=\mathbb{I}^{S}:\mathbf{\dot{\varepsilon}%
}-\mathbb{I}^{S}:\mathbf{\dot{\varepsilon}}_{p}=\mathbf{\dot{\varepsilon}%
}-\mathbf{\dot{\varepsilon}}_{p}=\left.  \mathbf{\dot{\varepsilon}}%
_{e}\right\vert _{%
\epspdo
}+\left.  \mathbf{\dot{\varepsilon}}_{e}\right\vert _{%
\epsdo
} \label{epsedot}%
\end{equation}
where $\mathbb{I}^{S}$ stands for the fourth-order (symmetric) identity tensor%
\begin{equation}
(\mathbb{I}^{S})_{ijkl}=\frac{1}{2}\left(  \delta_{ik}\delta_{jl}+\delta
_{il}\delta_{jk}\right)
\end{equation}
For further use, we define the following \emph{partial} contributions to the
elastic strain rate tensor%
\begin{equation}
\left.  \mathbf{\dot{\varepsilon}}_{e}\right\vert _{%
\epspdo
}=\left.  \frac{\partial\mathbf{\varepsilon}_{e}}{\partial\mathbf{\varepsilon
}}\right\vert _{%
\epspdo
}:\mathbf{\dot{\varepsilon}}=\mathbb{I}^{S}:\mathbf{\dot{\varepsilon}%
}=\mathbf{\dot{\varepsilon}} \label{epsedot with epspdot0}%
\end{equation}
and%
\begin{equation}
\left.  \mathbf{\dot{\varepsilon}}_{e}\right\vert _{%
\epsdo
}=\left.  \frac{\partial\mathbf{\varepsilon}_{e}}{\partial\mathbf{\varepsilon
}_{p}}\right\vert _{%
\epsdo
}:\mathbf{\dot{\varepsilon}}_{p}=-\mathbb{I}^{S}:\mathbf{\dot{\varepsilon}%
}_{p}=-\mathbf{\dot{\varepsilon}}_{p} \label{epsedot with epsdot0}%
\end{equation}

The stored energy in the device of Figure \ref{Figure - LeeSmall0.eps} is
given in terms of the internal elastic deformation, i.e. $\Psi=\Psi\left(
\mathbf{\varepsilon}_{e}\right)  $. The (non-negative) dissipation rate
$\mathcal{D}$ is calculated from the stress power $\mathcal{P}$ and the total
strain energy rate $\dot{\Psi}$ through%
\begin{equation}
\mathcal{D}=\mathcal{P}-\dot{\Psi}\geq0
\end{equation}
which can be written as%
\begin{equation}
\mathcal{D}=\mathbf{\sigma}:\mathbf{\dot{\varepsilon}}-\mathbf{\sigma}%
^{|e}:\mathbf{\dot{\varepsilon}}_{e}\geq0 \label{Dissipationeps}%
\end{equation}
where we have introduced the following notation for the total gradient---we
use the notation $d\left(  \cdot\right)  /d(\circ)$ for total differentiation%
\begin{equation}
\mathbf{\sigma}^{|e}:=\frac{d\Psi\left(  \mathbf{\varepsilon}_{e}\right)
}{d\mathbf{\varepsilon}_{e}} \label{sigext and sigint}%
\end{equation}

No dissipation takes place if we consider an isolated evolution of the
external, independent variable $\mathbf{\dot{\varepsilon}}\neq\mathbf{0}$
without internal variable evolution, i.e. with $\mathbf{\dot{\varepsilon}}%
_{p}=\mathbf{0}$. Then, from Eq. (\ref{epsedot}) $\mathbf{\dot{\varepsilon}%
}_{e}=\mathbf{\dot{\varepsilon}}_{e}|_{%
\epspdo
}=\mathbf{\dot{\varepsilon}}$ and Eq. (\ref{Dissipationeps}) reads%
\begin{equation}
\mathcal{D}=\mathbf{\sigma}:\mathbf{\dot{\varepsilon}}-\mathbf{\sigma}%
^{|e}:\left.  \mathbf{\dot{\varepsilon}}_{e}\right\vert _{%
\epspdo
}=\left(  \mathbf{\sigma}-\mathbf{\sigma}^{|e}:\left.  \frac{\partial
\mathbf{\varepsilon}_{e}}{\partial\mathbf{\varepsilon}}\right\vert _{%
\epspdo
}\right)  :\mathbf{\dot{\varepsilon}}=0\text{\quad if}\quad\mathbf{\dot
{\varepsilon}}_{p}=\mathbf{0} \label{DissipationElastic}%
\end{equation}
which yields%
\begin{equation}
\mathbf{\sigma}=\mathbf{\sigma}^{|e}:\left.  \frac{\partial\mathbf{\varepsilon
}_{e}}{\partial\mathbf{\varepsilon}}\right\vert _{%
\epspdo
}=\mathbf{\sigma}^{|e}:\mathbb{I}^{S}=\mathbf{\sigma}^{|e}
\label{sig=sigext+sigint}%
\end{equation}
where we recognize the following definition based on a chain rule
operation---note the abuse of notation $\Psi\left(  \mathbf{\varepsilon}%
_{e}\right)  =\Psi\left(  \mathbf{\varepsilon}_{e}\left(  \mathbf{\varepsilon
},\mathbf{\varepsilon}_{p}\right)  \right)  =\Psi\left(  \mathbf{\varepsilon
},\mathbf{\varepsilon}_{p}\right)  $; we keep the dependencies explicitly
stated when the distinction is needed%
\begin{align}
\mathbf{\sigma}  &  =\mathbf{\sigma}^{|e}:\left.  \frac{\partial
\mathbf{\varepsilon}_{e}}{\partial\mathbf{\varepsilon}}\right\vert _{%
\epspdo
}=\frac{d\Psi}{d\mathbf{\varepsilon}_{e}}:\left.  \frac{\partial
\mathbf{\varepsilon}_{e}}{\partial\mathbf{\varepsilon}}\right\vert _{%
\epspdo
}\nonumber\\
&  =\frac{d\Psi\left(  \mathbf{\varepsilon}_{e}\right)  }{d\mathbf{\varepsilon
}_{e}}:\frac{\partial\mathbf{\varepsilon}_{e}\left(  \mathbf{\varepsilon
},\mathbf{\varepsilon}_{p}\right)  }{\partial\mathbf{\varepsilon}}%
=\frac{\partial\Psi\left(  \mathbf{\varepsilon},\mathbf{\varepsilon}%
_{p}\right)  }{\partial\mathbf{\varepsilon}}\equiv\left.  \frac{\partial\Psi
}{\partial\mathbf{\varepsilon}}\right\vert _{%
\epspdo
}%
\end{align}
These definitions based on the concept of partial differentiation relate
internal variables with external ones from a purely kinematical standpoint and
will prove extremely useful in the finite deformation context, where they will
furnish the proper pull-back and push-forward operations between the different
configurations being defined.

Consider now an isolated variation of the other independent variable in the
problem, i.e. the case for which $\mathbf{\dot{\varepsilon}}=\mathbf{0}$ and
$\mathbf{\dot{\varepsilon}}_{p}\neq\mathbf{0}$, which note is a purely
internal (dissipative) evolution. Then from Eq. (\ref{epsedot}) $\mathbf{\dot
{\varepsilon}}_{e}=\mathbf{\dot{\varepsilon}}_{e}|_{%
\epsdo
}$. The dissipation inequality of Eq. (\ref{Dissipationeps}) must be positive
because plastic deformation is taking place%
\begin{equation}
\mathcal{D}=-\mathbf{\sigma}^{|e}:\left.  \mathbf{\dot{\varepsilon}}%
_{e}\right\vert _{%
\epsdo
}>0\text{\quad if}\quad\mathbf{\dot{\varepsilon}}_{p}\neq\mathbf{0}
\label{Dissipationepse}%
\end{equation}
We arrive at the same expression of Eq. (\ref{Dissipationepse}) if we consider
the most general case for which both independent variables are simultaneously
evolving, i.e. $\mathbf{\dot{\varepsilon}}\neq\mathbf{0}$ and $\mathbf{\dot
{\varepsilon}}_{p}\neq\mathbf{0}$. Hence note that both Eqs.
(\ref{DissipationElastic}) and (\ref{Dissipationepse}) hold if either
$\mathbf{\dot{\varepsilon}}=\mathbf{0}$ or $\mathbf{\dot{\varepsilon}}%
\neq\mathbf{0}$, so only the respective condition over $\mathbf{\dot
{\varepsilon}}_{p}$ is indicated in those equations. Since in the
infinitesimal framework of this section $\mathbf{\sigma}=\mathbf{\sigma}^{|e}$
and $\mathbf{\dot{\varepsilon}}_{e}|_{%
\epsdo
}=-\mathbf{\dot{\varepsilon}}_{p}$, recall Eqs. (\ref{sig=sigext+sigint}) and
(\ref{epsedot with epsdot0}), just in this case we can write Eq.
(\ref{Dissipationepse}) in its conventional form%
\begin{equation}
\mathcal{D}=\mathbf{\sigma}:\mathbf{\dot{\varepsilon}}_{p}>0\text{\quad
if}\quad\mathbf{\dot{\varepsilon}}_{p}\neq\mathbf{0} \label{Dissipationepsp}%
\end{equation}
i.e. the dissipation must be positive when the (six-dimensional) frictional
element in Figure \ref{Figure - LeeSmall0.eps} experiences slip.
Interestingly, Equations (\ref{Dissipationepse}) and (\ref{Dissipationepsp})
represent both the same physical concept, the former written in terms of the
\emph{partial} contribution $\mathbf{\dot{\varepsilon}}_{e}|_{%
\epsdo
}$ to the rate of the \emph{dependent} internal variable $\mathbf{\varepsilon
}_{e}\left(  \mathbf{\varepsilon},\mathbf{\varepsilon}_{p}\right)  $ and the
latter written in terms of the \emph{total} rate $\mathbf{\dot{\varepsilon}%
}_{p}$ of the \emph{independent} internal variable $\mathbf{\varepsilon}_{p}$.
However note that they present a clearly different interpretation which will
become relevant in the large strain framework.

\subsection{Local evolution equation in terms of $\mathbf{\dot{\varepsilon}%
}_{e}|_{%
\epsdo
}$}

Equation (\ref{Dissipationepse}) is automatically fulfilled if we choose the
following evolution equation for the internal strains $\mathbf{\varepsilon
}_{e}$%
\begin{equation}
-\left.  \mathbf{\dot{\varepsilon}}_{e}\right\vert _{%
\epsdo
}=\dot{\gamma}\frac{1}{k}\mathbb{N}:\mathbf{\sigma}^{|e}
\label{flow rule small}%
\end{equation}
which yields%
\begin{equation}
\mathcal{D}=\frac{\mathbf{\sigma}^{|e}:\mathbb{N}:\mathbf{\sigma}^{|e}}{k^{2}%
}k\dot{\gamma}>0\text{\quad if}\quad\mathbf{\dot{\varepsilon}}_{p}\neq0
\label{Dissipationsige}%
\end{equation}
where $\mathbb{N}$ is a fully symmetric positive definite fourth-order tensor,
$k>0$ is the characteristic yield stress of the internal frictional element of
Figure \ref{Figure - LeeSmall0.eps} and $\dot{\gamma}\geq0$ is the plastic
strain rate component which is power-conjugate of the stress-like variable
$k$, as we see just below. If the internal yield stress $k$ is constant, the
model describes the perfect plasticity case. If $k=k\left(  \gamma\right)  $
increases with an increment of the amount of plastic deformation$\ \gamma
=\int_{0}^{t}\dot{\gamma}dt$, namely $dk\left(  \gamma\right)  /d\gamma
=k^{\prime}\left(  \gamma\right)  >0$, the model may incorporate non-linear
isotropic hardening effects. We rephrase the dissipation Equation
(\ref{Dissipationsige}) as%
\begin{equation}
\mathcal{D}=\frac{1}{k^{2}}\left(  \mathbf{\sigma}^{|e}:\mathbb{N}%
:\mathbf{\sigma}^{|e}-k^{2}\right)  k\dot{\gamma}+k\dot{\gamma}>0\text{\quad
if}\quad\dot{\gamma}>0 \label{Dissipation general small}%
\end{equation}
then we immediately recognize the yield function $f(\mathbf{\sigma}^{|e},k)$
and the loading-unloading conditions%
\begin{equation}
\dot{\gamma}>0\text{\quad}\Rightarrow\quad f(\mathbf{\sigma}^{|e}%
,k)=\mathbf{\sigma}^{|e}:\mathbb{N}:\mathbf{\sigma}^{|e}-k^{2}=0
\label{Yield fcn small}%
\end{equation}
and%
\begin{equation}
f(\mathbf{\sigma}^{|e},k)=\mathbf{\sigma}^{|e}:\mathbb{N}:\mathbf{\sigma}%
^{|e}-k^{2}<0\text{\quad}\Rightarrow\quad\dot{\gamma}=0
\label{elastic response small}%
\end{equation}
so we obtain the plastic dissipation (if any) as given by the (scalar) flow
stress times the (scalar) frictional strain rate $\mathcal{D}=k\dot{\gamma
}\geq0$ for $\dot{\gamma}\geq0$.

Equation (\ref{flow rule small}) may be reinterpreted in terms of the yield
function gradient $\frac{1}{2}\nabla f=\mathbb{N}:\mathbf{\sigma}^{|e}$ to
give the following associative flow rule for the internal elastic strains
evolution%
\begin{equation}
-\left.  \mathbf{\dot{\varepsilon}}_{e}\right\vert _{%
\epsdo
}=\dot{\gamma}\frac{1}{k}\nabla\phi\label{flow rule small gradient}%
\end{equation}
where we have introduced the quadratic form
\begin{equation}
\phi(\mathbf{\sigma}^{|e})=\frac{1}{2}\mathbf{\sigma}^{|e}:\mathbb{N}%
:\mathbf{\sigma}^{|e} \label{phi pequenas}%
\end{equation}
for the matter of notation simplicity, so $f(\mathbf{\sigma}^{|e}%
,k)=2\phi(\mathbf{\sigma}^{|e})-k^{2}=0$ and $\frac{1}{2}\nabla f=\nabla\phi$.

\subsection{Local evolution equation in terms of $\mathbf{\dot{\varepsilon}%
}_{p}$}

Using the equivalences given in Eqs. (\ref{sig=sigext+sigint}) and
(\ref{epsedot with epsdot0}), the yield function of Eq. (\ref{Yield fcn small}%
)\ is given in terms of the (external) stress tensor $\mathbf{\sigma}$ as%
\begin{equation}
f(\mathbf{\sigma},k)=\mathbf{\sigma}:\mathbb{N}:\mathbf{\sigma}-k^{2}%
=2\phi\left(  \mathbf{\sigma}\right)  -k^{2}=0\text{\quad if}\quad\dot{\gamma
}>0
\end{equation}
and the associative flow rule of Eq. (\ref{flow rule small gradient}) adopts
the usual expression in terms of the (internal) plastic strain rate tensor
$\mathbf{\dot{\varepsilon}}_{p}$, cf. Eq. (2.5.6) of Ref.
\cite{SimoHughesBook} or Eq. (87) of Ref. \cite{MinCamMon}%
\begin{equation}
\mathbf{\dot{\varepsilon}}_{p}=\dot{\gamma}\frac{\nabla\phi}{\sqrt
{\mathbf{\sigma}:\mathbb{N}:\mathbf{\sigma}}} \label{flow rule small plastic}%
\end{equation}
As we discuss below, the interpretation given in Eq.
(\ref{flow rule small gradient}) greatly facilitates the extension of the
infinitesimal formulation to the finite strain context without modification.

\subsection{Description in terms of \emph{trial} and \emph{corrector} elastic
strain rates}

It is apparent from the foregoing results that, in practice, no distinction is
needed within the infinitesimal framework regarding both the selection of
either $\mathbf{\varepsilon}_{e}$ or $\mathbf{\varepsilon}_{p}$ as the basic
internal deformation variable and the selection of either $\mathbf{\sigma
}^{|e}$ or $\mathbf{\sigma}$ as the basic stress tensor. In what follows,
however, we keep on developing the infinitesimal formulation in terms of
$\mathbf{\varepsilon}_{e}$ and $\mathbf{\sigma}^{|e}$, which will let us take
special advantage of the functional dependencies $\mathbf{\varepsilon}%
_{e}\left(  \mathbf{\varepsilon},\mathbf{\varepsilon}_{p}\right)
=\mathbf{\varepsilon}-\mathbf{\varepsilon}_{p}$ and $\mathbf{\sigma}%
^{|e}\left(  \mathbf{\varepsilon}_{e}\right)  =d\Psi\left(
\mathbf{\varepsilon}_{e}\right)  /d\mathbf{\varepsilon}_{e}$.

Regarding the evolution of elastic variables, whether strains or stresses, it
is convenient to introduce the concepts of \emph{trial} and \emph{corrector}
elastic strain rates in Eq. (\ref{epsedot}). This decomposition in rate form
is the origin of the trial elastic predictor, for which $\mathbf{\varepsilon
}_{p}$ is frozen, and plastic corrector, for which $\mathbf{\varepsilon}$ is
frozen, operator split typically employed for elastic internal variables in
computational inelasticity within an algorithmic framework. Accordingly, we
define within the \emph{continuum} theory%
\begin{equation}
\mathbf{\dot{\varepsilon}}_{e}=\left.  \mathbf{\dot{\varepsilon}}%
_{e}\right\vert _{%
\epspdo
}+\left.  \mathbf{\dot{\varepsilon}}_{e}\right\vert _{%
\epsdo
}=:\,^{tr}\mathbf{\dot{\varepsilon}}_{e}+\,^{ct}\mathbf{\dot{\varepsilon}}_{e}
\label{epsedot tr ct}%
\end{equation}
where the superscripts \emph{tr} and \emph{ct} stand for \emph{trial} and
\emph{corrector} respectively. Interestingly, the concepts of \emph{trial} and
\emph{corrector} elastic rates emerge in the finite deformation multiplicative
framework developed below without modification with respect to the
infinitesimal case, so we will be able to directly compare the small and large
strain formulations equation by equation. We note that elastoplasticity models
based on plastic metrics have traditionally followed the same philosophy, but
departing from the standard rate decomposition%
\begin{equation}
\mathbf{\dot{\varepsilon}}_{e}=\mathbf{\dot{\varepsilon}}-\mathbf{\dot
{\varepsilon}}_{p}%
\end{equation}
which, however, leads to additive Lagrangian formulations \cite{GreenNaghdi65}%
, \cite{Miehe98}, \cite{PapadopoulusLu98}, \cite{PapadopoulusLu01},
\cite{Miehe02}, \cite{LobleinSchroderGruttmann03}, \cite{SansourWagner03},
\cite{Ulz09} that are not generally consistent with the finite strain
multiplicative decomposition, as it is well-known \cite{GreenNaghdi71},
\cite{Itskov04}, \cite{Schmidt05}, \cite{NeffGhiba16}.

For further comparison, we rephrase both the dissipation inequality of Eq.
(\ref{Dissipationepse}) and the associative flow rule of Eq.
(\ref{flow rule small gradient}) in terms of the corrector elastic strain rate
as%
\begin{equation}
\mathcal{D}=-\mathbf{\sigma}^{|e}:\,^{ct}\mathbf{\dot{\varepsilon}}%
_{e}>0\text{\quad if}\quad\dot{\gamma}>0 \label{Dissipation epse ct}%
\end{equation}
and%
\begin{equation}
\,^{ct}\mathbf{\dot{\varepsilon}}_{e}=-\dot{\gamma}\frac{1}{k}\nabla
\phi\label{flow rule small gradient ct}%
\end{equation}
Note that the elastic strain correction performed in CPP algorithms and
defined in Eq. (\ref{flow rule small gradient ct}) enforce the instantaneous
closest point projection onto the elastic domain, i.e. the normality rule in
the continuum setting.

In the case we do not consider a potential, then the formulation is usually
referred to as generalized\ plasticity \cite{PastorZ}, which is a
generalization of nonassociative plasticity typically used in soils
\cite{Borjabook}. However, we can alternatively take%
\begin{equation}
\,^{ct}\mathbf{\dot{\varepsilon}}_{e}=-\dot{\gamma}\frac{1}{k}\mathbf{G}%
(\mathbf{\sigma}^{|e}) \label{flow rule nonassociative}%
\end{equation}
where the prescribed second-order tensor function $\mathbf{G}(\mathbf{\sigma
}^{|e})$ defines the direction of plastic flow. So Eq.
(\ref{Dissipation epse ct}) reads%
\begin{equation}
\mathcal{D}=\dot{\gamma}\frac{1}{k}\mathbf{\sigma}^{|e}:\mathbf{G}%
\quad\text{if}\quad\dot{\gamma}>0 \label{dissipation nonassociative}%
\end{equation}
even though positive dissipation and a fully symmetric linearization of the
continuum theory are not guaranteed in this case \cite{Simo98}. Note that
$\mathbf{G}=\nabla\phi$ for associative plasticity.

\subsection{Maximum Plastic Dissipation}

We assume now the existence of another arbitrary stress field $\mathbf{\Sigma
}=\mathbf{\Sigma}^{|e}$ different from the actual stress field $\mathbf{\sigma
}=\mathbf{\sigma}^{|e}$, as given in\ Eq. (\ref{sig=sigext+sigint}). The
dissipation originated by $\mathbf{\Sigma}^{|e}$ during the same plastic flow
process would be---cf. Eq. (\ref{Dissipation epse ct})%
\begin{equation}
\mathcal{D}_{\Sigma}=-\mathbf{\Sigma}^{|e}:\,^{ct}\mathbf{\dot{\varepsilon}%
}_{e}\text{\quad if}\quad\dot{\gamma}>0
\end{equation}
The evolution of plastic flow, e.g. Eq. (\ref{flow rule small gradient ct}),
is said to obey the Principle of Maximum Dissipation if%
\begin{equation}
\mathcal{D}-\mathcal{D}_{\Sigma}>0
\end{equation}
for any admissible stress field $\mathbf{\Sigma}^{|e}\neq\mathbf{\sigma}^{|e}%
$, i.e. with $f(\mathbf{\Sigma}^{|e},k)\leq0$. Considering the associative
flow rule of Eq. (\ref{flow rule small gradient ct}), we arrive at%
\begin{equation}
\mathcal{D}-\mathcal{D}_{\Sigma}=-(\mathbf{\sigma}^{|e}-\mathbf{\Sigma}%
^{|e}):\,^{ct}\mathbf{\dot{\varepsilon}}_{e}=\dot{\gamma}\frac{1}%
{k}(\mathbf{\sigma}^{|e}-\mathbf{\Sigma}^{|e}):\nabla\phi
\end{equation}
If $f(\mathbf{\sigma}^{|e},k)=2\phi(\mathbf{\sigma}^{|e})-k^{2}=0$ is a
strictly convex function and $\mathbf{\Sigma}^{|e}$ is admissible%
\begin{equation}
\mathcal{D}-\mathcal{D}_{\Sigma}=\dot{\gamma}\frac{1}{k}(\mathbf{\sigma}%
^{|e}-\mathbf{\Sigma}^{|e}):\nabla\phi=\dot{\gamma}\frac{1}{2k}(\mathbf{\sigma
}^{|e}-\mathbf{\Sigma}^{|e}):\nabla f>0
\end{equation}
i.e. maximum dissipation in the system is guaranteed (the equal sign would be
possible if non-strictly convex functions are considered, as for example
Tresca's one).

In all the finite strain cases addressed below $\mathcal{D}-\mathcal{D}%
_{\Sigma}>0$ if the corresponding associative flow rule for each case is
considered. Indeed, this principle must hold in any arbitrary stress-strain
work-conjugate couple, but if guaranteed in one of them, will hold in any of
them by invariance of power.

\begin{table}[h]
\caption{Small strain additive anisotropic elastoplasticity model.}%
\label{Table - Small}
\begin{center}%
\begin{tabular}
[c]{|l|}\hline%
\begin{tabular}
[c]{l}%
\vspace{-0.1in}\\
(i) Additive decomposition of the strain $\mathbf{\varepsilon}%
=\mathbf{\varepsilon}_{e}+\mathbf{\varepsilon}_{p}$\bigskip\\
(ii) Symmetric internal strain variable $\mathbf{\varepsilon}_{e}$\bigskip\\
(iii) Kinematics induced by $\mathbf{\varepsilon}_{e}(\mathbf{\varepsilon
},\mathbf{\varepsilon}_{p})=\mathbf{\varepsilon}-\mathbf{\varepsilon}_{p}%
$\medskip\\
\qquad$\mathbf{\dot{\varepsilon}}_{e}=\left.  \mathbf{\dot{\varepsilon}}%
_{e}\right\vert _{{\scriptsize \mathbf{\dot{\varepsilon}}_{p}=\mathbf{0}}%
}+\left.  \mathbf{\dot{\varepsilon}}_{e}\right\vert
_{{\scriptsize \mathbf{\dot{\varepsilon}}=\mathbf{0}}}=\,^{tr}\mathbf{\dot
{\varepsilon}}_{e}+\,^{ct}\mathbf{\dot{\varepsilon}}_{e}\equiv\mathbf{\dot
{\varepsilon}}-\mathbf{\dot{\varepsilon}}_{p}$\bigskip\\
(iv) Symmetric stresses deriving from the strain energy $\Psi
(\mathbf{\varepsilon}_{e})$\medskip\\
\qquad$\mathbf{\sigma}^{|e}=\dfrac{d\Psi(\mathbf{\varepsilon}_{e}%
)}{d\mathbf{\varepsilon}_{e}}$\,,\qquad$\mathbf{\sigma}=\dfrac{\partial
\Psi(\mathbf{\varepsilon},\mathbf{\varepsilon}_{p})}{\partial
\mathbf{\varepsilon}}=\mathbf{\sigma}^{|e}:\dfrac{\partial\mathbf{\varepsilon
}_{e}(\mathbf{\varepsilon},\mathbf{\varepsilon}_{p})}{\partial
\mathbf{\varepsilon}}\equiv\mathbf{\sigma}^{|e}$\bigskip\\
(v) Evolution equation for associative symmetric plastic flow\medskip\\
\qquad$-\,^{ct}\mathbf{\dot{\varepsilon}}_{e}=\dot{\gamma}\dfrac{1}{k}%
\nabla\phi(\mathbf{\sigma}^{|e})\equiv\mathbf{\dot{\varepsilon}}_{p}$%
\medskip\\
\qquad$\dot{\gamma}\geq0$\,,\quad$f(\mathbf{\sigma}^{|e},k)=2\phi
(\mathbf{\sigma}^{|e})-k^{2}\leq0$\,,\quad$\dot{\gamma}f(\mathbf{\sigma}%
^{|e},k)=0$\bigskip\\
{\small {Note: Potential $\Psi(\mathbf{\varepsilon}_{e})$ and function
$f(\mathbf{\sigma}^{|e},k)$ are anisotropic, in general.}\bigskip}\\
\end{tabular}
\\\hline
\end{tabular}
\end{center}
\end{table}

\section{Finite strain anisotropic elastoplasticity formulated in the
current\ configuration}

We present in this section a new framework for finite strain anisotropic
elastoplasticity formulated in the current configuration in which the basic
internal variables are elastic in nature. Once the corresponding dependencies
are identified, the theory is further developed taking advantage of the
previously introduced concepts of partial differentiation, mapping tensors and
the trial-corrector decomposition of internal elastic variables in rate form.
With the exception of the geometrical nonlinearities being introduced, the
formulation yields identical expressions to those derived above for
infinitesimal plasticity.

\subsection{Multiplicative decomposition}

The so-called Lee multiplicative decomposition \cite{Lee69}\ states the
decomposition of the deformation gradient into an elastic part and a plastic
part as%
\begin{equation}
\mathbf{X}=\mathbf{X}_{e}\mathbf{X}_{p} \label{LeeDecomposition}%
\end{equation}
When using this decomposition, a superimposed rigid body motion by an
orthogonal proper tensor $\mathbf{Q}$ results into%
\begin{equation}
\mathbf{X}^{+}=\mathbf{QX}=\mathbf{X}_{e}^{+}\mathbf{X}_{p}^{+}=\left(
\mathbf{QX}_{e}\right)  \left(  \mathbf{X}_{p}\right)  \label{rigidbodymotion}%
\end{equation}
so the rigid body motion naturally enters the \textquotedblleft
elastic\textquotedblright\ gradient, whereas the plastic gradient remains
unaltered. A much debated issue is the uniqueness of the intermediate
configuration arising from $\mathbf{X}_{p}$ since any arbitrary rotation
tensor $\mathbf{Q}$ with its inverse may be inserted such that $\mathbf{X}%
=\left(  \mathbf{X}_{e}\mathbf{Q}\right)  (\mathbf{Q}^{T}\mathbf{X}_{p})$, so
the decomposition of Eq. (\ref{LeeDecomposition}) is unique up to a rigid body
rotation of the intermediate configuration. However, in practice, since
$\mathbf{X}_{p}$ \emph{is path dependent} and is integrated step-by-step in an
incremental fashion in computational elastoplasticity algorithms
\cite{CamineroMontansBathe11,MontansBenitezCaminero12}, we consider that it is
uniquely determined at all times.

\subsection{\emph{Trial} and \emph{corrector} elastic deformation rate
tensors}

Consider the following additive decomposition of the spatial velocity gradient
tensor%
\begin{equation}
\mathbf{l}:=\mathbf{\dot{X}X}^{-1}=\mathbf{\dot{X}}_{e}\mathbf{X}_{e}%
^{-1}+\mathbf{X}_{e}\mathbf{\dot{X}}_{p}\mathbf{X}_{p}^{-1}\mathbf{X}_{e}%
^{-1}=\mathbf{l}_{e}+\mathbf{X}_{e}\mathbf{l}_{p}\mathbf{X}_{e}^{-1}
\label{l=le+lp}%
\end{equation}
where we define the elastic and plastic velocity gradients as%
\begin{equation}
\mathbf{l}_{e}:=\mathbf{\dot{X}}_{e}\mathbf{X}_{e}^{-1}\text{\quad and\quad
}\mathbf{l}_{p}:=\mathbf{\dot{X}}_{p}\mathbf{X}_{p}^{-1}%
\end{equation}
We note that $\mathbf{l}_{e}$ lies in the spatial configuration, whereas
$\mathbf{l}_{p}$ operates in the intermediate configuration. The deformation
rate tensor (the symmetric part of $\mathbf{l}$) and the spin tensor (its
skew-symmetric part) are%
\begin{equation}
\mathbf{d}=sym\left(  \mathbf{l}\right)  \text{\quad and\quad}\mathbf{w}%
=skw\left(  \mathbf{l}\right)
\end{equation}
The elastic and plastic velocity gradient tensors also admit the corresponding
decomposition into deformation rate and spin counterparts, $\mathbf{l}%
_{e}=\mathbf{d}_{e}+\mathbf{w}_{e}$ and $\mathbf{l}_{p}=\mathbf{d}%
_{p}+\mathbf{w}_{p}$, thereby from Eq. (\ref{l=le+lp})%
\begin{equation}
\mathbf{d}=\mathbf{d}_{e}+sym\left(  \mathbf{X}_{e}\mathbf{l}_{p}%
\mathbf{X}_{e}^{-1}\right)  \label{d=de+lp}%
\end{equation}%
\begin{equation}
\mathbf{w}=\mathbf{w}_{e}+skw\left(  \mathbf{X}_{e}\mathbf{l}_{p}%
\mathbf{X}_{e}^{-1}\right)
\end{equation}

In general, from Eq. (\ref{d=de+lp}) we can consider the elastic deformation
rate tensor as a two-variable function of the deformation rate tensor and the
plastic velocity gradient tensor (including the plastic spin $\mathbf{w}_{p}$)
through%
\begin{equation}
\mathbf{d}_{e}(\mathbf{d,l}_{p})=\mathbf{d}-sym\left(  \mathbf{X}%
_{e}\mathbf{l}_{p}\mathbf{X}_{e}^{-1}\right)  \label{de d lp}%
\end{equation}
which can be expressed in the following rate-form formats---compare with Eqs.
(\ref{epsedot}) and (\ref{epsedot tr ct})%
\begin{equation}
\mathbf{d}_{e}=\left.  \mathbb{M}_{d}^{d_{e}}\right\vert _{%
\lpo
}:\mathbf{d}+\left.  \mathbb{M}_{l_{p}}^{d_{e}}\right\vert _{%
\doo
}:\mathbf{l}_{p}=\left.  \mathbf{d}_{e}\right\vert _{%
\lpo
}+\left.  \mathbf{d}_{e}\right\vert _{%
\doo
}=\,^{tr}\mathbf{d}_{e}+\,^{ct}\mathbf{d}_{e} \label{de tr ct}%
\end{equation}
where $\left.  \mathbb{M}_{d}^{d_{e}}\right\vert _{%
\lpo
}$ and $\left.  \mathbb{M}_{l_{p}}^{d_{e}}\right\vert _{%
\doo
}$ are mapping tensors \cite{CamineroMontansBathe11,LatMonAPM2016} which allow
us to define the following \emph{partial} contributions to the elastic
deformation rate tensor $\mathbf{d}_{e}$%
\begin{equation}
\,^{tr}\mathbf{d}_{e}:=\left.  \mathbb{M}_{d}^{d_{e}}\right\vert _{%
\lpo
}:\mathbf{d}=\mathbb{I}^{S}:\mathbf{d}=\mathbf{d} \label{de tr}%
\end{equation}
and%
\begin{equation}
^{ct}\mathbf{d}_{e}=\left.  \mathbb{M}_{l_{p}}^{d_{e}}\right\vert _{%
\doo
}:\mathbf{l}_{p}=-\frac{1}{2}\left(  \mathbf{X}_{e}\odot\mathbf{X}_{e}%
^{-T}+\mathbf{X}_{e}^{-T}\boxdot\mathbf{X}_{e}\right)  :\mathbf{l}%
_{p}=-sym\left(  \mathbf{X}_{e}\mathbf{l}_{p}\mathbf{X}_{e}^{-1}\right)
\label{de ct}%
\end{equation}
with $(\mathbf{Y}\odot\mathbf{Z})_{ijkl}=Y_{ik}Z_{jl}$ and $(\mathbf{Y\boxdot
Z})_{ijkl}=Y_{il}Z_{jk}$.

It is frequently assumed in computational plasticity that the plastic spin
vanishes, namely $\mathbf{w}_{p}=\mathbf{0}$, so its effects in the
dissipation inequality are not taken into account. However, as in the small
strain case discussed above, the plastic spin evolves independently of the
normality flow rules being developed below in terms of corrector elastic
rates, so no additional\ assumptions over $\mathbf{w}_{p}$ will be prescribed
by the dissipation process \cite{Lubliner86}. The a priori undetermined
intermediate configuration, defined by $\mathbf{X}_{p}$, would become
determined once an independent constitutive equation for the plastic spin
$\mathbf{w}_{p}$ is specified \cite{Simo98}, \cite{MontansBathe07},
\cite{KimMontansBathe09}, which is strictly needed in order to complete the
model formulation.

\subsection{Dissipation inequality and flow rule in terms of $\,^{ct}%
\mathbf{d}_{e}$}

From purely physical grounds, we know that the strain energy function locally
depends on an elastic measure of the deformation. Hence it may be expressed in
terms of a Lagrangian-like elastic strain tensor lying in the intermediate
configuration, e.g. the elastic Green--Lagrange-like strains $\mathbf{A}%
_{e}=\frac{1}{2}(\mathbf{X}_{e}^{T}\mathbf{X}_{e}-\mathbf{I})$ where
$\mathbf{I}$ is the second-order identity tensor, as%
\begin{equation}
\Psi_{A}=\Psi_{A}\left(  \mathbf{A}_{e},\mathbf{a}_{1}\otimes\mathbf{a}%
_{1},\mathbf{a}_{2}\otimes\mathbf{a}_{2}\right)  \label{Psi(Ae)}%
\end{equation}
where we have additionally assumed that the material is orthotropic, with
$\mathbf{a}_{1}$ and $\mathbf{a}_{2}$ (and $\mathbf{a}_{3}=\mathbf{a}%
_{1}\times\mathbf{a}_{2}$) defining the orthogonal preferred directions in the
intermediate configuration. As a first step in the derivation of more complex
formulations including texture evolution, which involves an experimentally
motivated constitutive equation additional to that for $\mathbf{w}_{p}$, see
examples in Ref. \cite{KimMontansBathe09} and references therein, we assume in
this work that the texture of the material is permanent and independent of the
plastic spin. That is, we consider the case for which $\mathbf{w}_{p}%
\neq\mathbf{0}$ is given as an additional equation so that the Lee
decomposition is completely defined at each instant and we take $\mathbf{\dot
{a}}_{1}=\mathbf{\dot{a}}_{2}=\mathbf{\dot{a}}_{3}=\mathbf{0}$ as a
simplifying assumption for the stresses update. Subsequently, the material
time derivative of the Lagrangian potential $\Psi_{A}$ may be expressed in
terms of variables lying in the current configuration through%
\begin{equation}
\dot{\Psi}_{A}=\frac{d\Psi_{A}\left(  \mathbf{A}_{e}\right)  }{d\mathbf{A}%
_{e}}:\mathbf{\dot{A}}_{e}=\mathbf{S}^{|e}:\mathbf{\dot{A}}_{e}=\mathbf{S}%
^{|e}:\mathbf{X}_{e}^{T}\odot\mathbf{X}_{e}^{T}:\mathbf{d}_{e}=\mathbf{\tau
}^{|e}:\mathbf{d}_{e} \label{Psidot Aedot de}%
\end{equation}
where we have used the purely kinematical pull-back operation over
$\mathbf{d}_{e}$ (lying in the current configuration)\ that gives
$\mathbf{\dot{A}}_{e}$ (lying in the intermediate configuration) ---see
\cite{LatMonAPM2016}%
\begin{equation}
\mathbf{\dot{A}}_{e}=\mathbf{X}_{e}^{T}\mathbf{d}_{e}\mathbf{X}_{e}%
=\mathbf{X}_{e}^{T}\odot\mathbf{X}_{e}^{T}:\mathbf{d}_{e}=:\mathbb{M}_{d_{e}%
}^{\dot{A}_{e}}:\mathbf{d}_{e}%
\end{equation}
which provides as a result the also purely kinematical push-forward operation
over the internal elastic second Piola--Kirchhoff stress tensor (lying in the
intermediate configuration)%
\begin{equation}
\mathbf{S}^{|e}:=\frac{d\Psi_{A}\left(  \mathbf{A}_{e}\right)  }%
{d\mathbf{A}_{e}} \label{S|e}%
\end{equation}
that gives the internal elastic Kirchhoff stress tensor $\mathbf{\tau}^{|e}$
(lying in the current configuration)%
\begin{equation}
\mathbf{\tau}^{|e}:=\mathbf{S}^{|e}:\mathbb{M}_{d_{e}}^{\dot{A}_{e}%
}=\mathbf{S}^{|e}:\mathbf{X}_{e}^{T}\odot\mathbf{X}_{e}^{T}=\mathbf{X}%
_{e}\mathbf{S}^{|e}\mathbf{X}_{e}^{T} \label{taue Se}%
\end{equation}

For further use, we can define the elastic Kirchhoff stress tensor
$\mathbf{\tau}^{|e}$ from the elastic Almansi strain tensor $\mathbf{a}%
_{e}:=\frac{1}{2}(\mathbf{I}-\mathbf{X}_{e}^{-T}\mathbf{X}_{e}^{-1})$, both
operating in the current placement, through partial differentiation of the
strain energy function expressed in terms of the corresponding spatial
variables. To this end, we first recall from scratch that \emph{different}
strain tensors, whether material or spatial, are referential (intensive)
variables in the sense that they give local measures of the \emph{same}
(extensive) deformation with respect to \emph{different} reference line
elements. For example, consider the following (contravariant) relation between
the elastic Almansi strain tensor $\mathbf{a}_{e}$ and the elastic
Green--Lagrange-like one $\mathbf{A}_{e}$%
\begin{equation}
\mathbf{a}_{e}(\mathbf{A}_{e};\mathbf{X}_{e})=\mathbf{X}_{e}^{-T}%
\mathbf{A}_{e}\mathbf{X}_{e}^{-1}=\mathbf{X}_{e}^{-T}\odot\mathbf{X}_{e}%
^{-T}:\mathbf{A}_{e}=\frac{\partial\mathbf{a}_{e}\left(  \mathbf{A}%
_{e};\mathbf{X}_{e}\right)  }{\partial\mathbf{A}_{e}}:\mathbf{A}_{e}
\label{ae(Ae;Xe)}%
\end{equation}
where we have intentionally separated the tensor variable dependencies
$\mathbf{a}_{e}\left(  \mathbf{A}_{e};\mathbf{X}_{e}\right)  $ with a
semicolon in order to make explicit the clearly different nature of both
dependencies; the left-hand argument includes information about the
\emph{same} elastic deformation process that $\mathbf{a}_{e}$ and
$\mathbf{A}_{e}$ are measuring; the right-hand argument just includes
information about the \emph{different} referential configuration to which
$\mathbf{a}_{e}$ and $\mathbf{A}_{e}$ are being referred. We want to remark
the conceptual difference existing between the functional dependence
$\mathbf{a}_{e}(\mathbf{A}_{e};\mathbf{X}_{e})$ in Eq. (\ref{ae(Ae;Xe)}),
which includes information about a \emph{single} deformation process (hence we
use a semicolon), with the functional dependence $\mathbf{\varepsilon}%
_{e}\left(  \mathbf{\varepsilon},\mathbf{\varepsilon}_{p}\right)  $ in Eq.
(\ref{epse dependence 0}), which includes information about two
\emph{different} deformation processes (hence we use a comma). As it is well
known, the material derivative of $\mathbf{a}_{e}$ is%
\begin{equation}
\mathbf{a}_{e}=\mathbf{X}_{e}^{-T}\mathbf{A}_{e}\mathbf{X}_{e}^{-1}%
\;\Rightarrow\;\mathbf{\dot{a}}_{e}=\overset{\diamond}{\mathbf{a}}%
_{e}-\mathbf{l}_{e}^{T}\mathbf{a}_{e}-\mathbf{a}_{e}\mathbf{l}_{e}%
^{T}=:\overset{\diamond}{\mathbf{a}}_{e}+\,\mathbf{\breve{a}}_{e}%
\end{equation}
where%
\begin{equation}
\overset{\diamond}{\mathbf{a}}_{e}=\mathbf{X}_{e}^{-T}\mathbf{\dot{A}}%
_{e}\mathbf{X}_{e}^{-1}\equiv\mathbf{d}_{e}\equiv\mathcal{L}_{e}\left(
\mathbf{a}_{e}\right)
\end{equation}
is the Lie (or Oldroyd) derivative of $\mathbf{a}_{e}$, and$\,\mathbf{\breve
{a}}_{e}$ are the convective ones. The material time derivative of
$\mathbf{a}_{e}$ may also be derived in a better form for interpretation, as
given in Eq. (\ref{ae(Ae;Xe)})%
\begin{equation}
\mathbf{\dot{a}}_{e}=\frac{\partial\mathbf{a}_{e}\left(  \mathbf{A}%
_{e};\mathbf{X}_{e}\right)  }{\partial\mathbf{A}_{e}}:\mathbf{\dot{A}}%
_{e}+\frac{\partial\mathbf{a}_{e}\left(  \mathbf{A}_{e};\mathbf{X}_{e}\right)
}{\partial\mathbf{X}_{e}}:\mathbf{\dot{X}}_{e}=\overset{\diamond}{\mathbf{a}%
}_{e}+\,\mathbf{\breve{a}}_{e} \label{aedot total}%
\end{equation}
so we can also interpret $\overset{\diamond}{\mathbf{a}}_{e}=\mathcal{L}%
_{e}\left(  \mathbf{a}_{e}\right)  $ through partial differentiation as%
\begin{equation}
\overset{\diamond}{\mathbf{a}}_{e}=\frac{\partial\mathbf{a}_{e}\left(
\mathbf{A}_{e};\mathbf{X}_{e}\right)  }{\partial\mathbf{A}_{e}}:\mathbf{\dot
{A}}_{e}=\mathbf{X}_{e}^{-T}\odot\mathbf{X}_{e}^{-T}:\mathbf{\dot{A}}%
_{e}=\mathbf{d}_{e} \label{aedot objective}%
\end{equation}
We can observe in\ Eq. (\ref{aedot total}) that, for a given local elastic
deformation state defined by $\mathbf{A}_{e}$ and $\mathbf{X}_{e}$, the
contribution $\overset{\diamond}{\mathbf{a}}_{e}\equiv\mathbf{d}_{e}$ to the
total rate $\mathbf{\dot{a}}_{e}$ depends on the \emph{objective} material
strain rate tensor $\mathbf{\dot{A}}_{e}$ only (i.e. a \textquotedblleft
true\textquotedblright\ deformation rate keeping the\ spatial reference fixed)
and that the contribution $\mathbf{\breve{a}}_{e}$ to the total rate
$\mathbf{\dot{a}}_{e}$ depends on the \emph{non-objective} deformation rate
tensor $\mathbf{\dot{X}}_{e}$ only (i.e. a true spatial reference
configuration rate keeping the deformation fixed). The latter contribution
gives rise, indeed, to the well-known convective terms resulting in lack of
objectivity of spatial variable rates.

As also well-known, the Lie (Oldroyd) derivative of $\mathbf{\tau}^{|e}$ is%
\begin{equation}
\overset{\diamond}{\mathbf{\tau}}^{|e}=\mathbf{X}_{e}\mathbf{\dot{S}}%
^{|e}\mathbf{X}_{e}^{T}\equiv\mathcal{L}_{e}\left(  \mathbf{\tau}^{|e}\right)
\end{equation}
Consider now the dependencies $\mathbf{\tau}^{|e}(\mathbf{S}^{|e}%
;\mathbf{X}_{e})$. The rate of change of $\mathbf{\tau}^{|e}$ with its spatial
reference being fixed may be written in a better form for interpretation as%
\begin{equation}
\overset{\diamond}{\mathbf{\tau}}^{|e}=\mathbf{X}_{e}\odot\mathbf{X}%
_{e}:\mathbf{\dot{S}}^{|e}=\frac{\partial\mathbf{\tau}^{|e}\left(
\mathbf{S}^{|e};\mathbf{X}_{e}\right)  }{\partial\mathbf{S}^{|e}}%
:\mathbf{\dot{S}}^{|e}%
\end{equation}
The previous lines emphasize that the terms $\overset{\diamond}{\mathbf{a}%
}_{e}$ and $\overset{\diamond}{\mathbf{\tau}}^{|e}$ are the relevant
derivatives to be used in the constitutive equations because they contain
respectively the partial derivatives of the respective spatial measures
$\mathbf{a}_{e}$ and $\mathbf{\tau}^{|e}$ respect to the change of the
quantities $\mathbf{A}_{e}$ and $\mathbf{S}^{|e}$ in the invariant reference configuration.

The interpretation given to $\mathbf{a}_{e}\left(  \mathbf{A}_{e}%
;\mathbf{X}_{e}\right)  $ allow us to define the elastic Kirchhoff stress
tensor $\mathbf{\tau}^{|e}$ from the elastic Almansi strain tensor
$\mathbf{a}_{e}$ via the Eulerian description of the strain energy function
$\Psi_{a}$, as we show next. Since%
\begin{equation}
\Psi_{A}\left(  \mathbf{A}_{e}\right)  =\Psi_{a}\left(  \mathbf{a}%
_{e};\mathbf{X}_{e}\right)  =\Psi_{a}\left(  \mathbf{a}_{e}\left(
\mathbf{A}_{e};\mathbf{X}_{e}\right)  ;\mathbf{X}_{e}\right)
\end{equation}
we have%
\begin{equation}
\dot{\Psi}_{A}\left(  \mathbf{A}_{e}\right)  =\mathbf{S}^{|e}:\mathbf{\dot{A}%
}_{e}=\frac{d\Psi_{A}}{d\mathbf{A}_{e}}:\mathbf{\dot{A}}_{e}=\frac
{\partial\Psi_{a}\left(  \mathbf{a}_{e};\mathbf{X}_{e}\right)  }%
{\partial\mathbf{a}_{e}}:\overset{\diamond}{\mathbf{a}}_{e}=\mathbf{\tau}%
^{|e}:\mathbf{d}_{e}=\,\overset{\diamond}{\Psi}_{a}\left(  \mathbf{a}%
_{e};\mathbf{X}_{e}\right)  \label{Psi def}%
\end{equation}
and we obtain $\mathbf{\tau}^{|e}$ from $\mathbf{a}_{e}$ based on the concept
of partial differentiation---see Ref. \cite{LatMonAPM2016} for an equivalent
result in terms of $\mathbf{\tau}$ and $\mathbf{a}$%

\begin{equation}
\mathbf{\tau}^{|e}=\frac{\partial\Psi_{a}\left(  \mathbf{a}_{e};\mathbf{X}%
_{e}\right)  }{\partial\mathbf{a}_{e}} \label{tauext and tauint}%
\end{equation}
where we would need to know the explicit dependence of $\Psi_{a}$ on both
$\mathbf{a}_{e}$ and $\mathbf{X}_{e}$. We observe in Eq. (\ref{Psi def}) that
both $\dot{\Psi}_{A}$ and $\overset{\diamond}{\Psi}_{a}$ represent the change
of the elastic potential $\Psi$ associated to true (i.e. objective) strain
rates, whether material or spatial.

Using Eq. (\ref{Psidot Aedot de}) and the stress power density per unit
reference volume $\mathcal{P}=\mathbf{\tau}:\mathbf{d}$, the dissipation
inequality written in the current configuration reads%

\begin{equation}
\mathcal{D}=\mathcal{P}-\dot{\Psi}_{A}=\mathcal{P}-\overset{\diamond}{\Psi
}_{a}=\mathbf{\tau}:\mathbf{d}-\mathbf{\tau}^{|e}:\mathbf{d}_{e}\geq0
\label{Dissipation spatial general}%
\end{equation}
where $\mathbf{\tau}$ is the Kirchhoff stress tensor, power-conjugate of the
deformation rate tensor $\mathbf{d}$ \cite{LatMonAPM2016}. Using the
decomposition given in Eq. (\ref{de tr ct}), Eq.
(\ref{Dissipation spatial general}) can be written as%
\begin{equation}
\mathcal{D}=\mathbf{\tau}:\mathbf{d}-\mathbf{\tau}^{|e}:\left(  \,^{tr}%
\mathbf{d}_{e}+\,^{ct}\mathbf{d}_{e}\right)  \geq0 \label{Dissipationd}%
\end{equation}
For the case of $\mathbf{l}_{p}=\mathbf{0}$, i.e. $\mathbf{d}_{e}%
=\,^{tr}\mathbf{d}_{e}$, we have no dissipation%
\begin{equation}
\mathcal{D}=\mathbf{\tau}:\mathbf{d}-\mathbf{\tau}^{|e}:\,^{tr}\mathbf{d}%
_{e}=\left(  \mathbf{\tau}-\mathbf{\tau}^{|e}:\left.  \mathbb{M}_{d}^{d_{e}%
}\right\vert _{%
\lpo
}\right)  :\mathbf{d}=0\quad\text{if}\quad\mathbf{l}_{p}=\mathbf{0}%
\end{equation}
so we obtain the following definition of the external Kirchhoff stresses
$\mathbf{\tau}$ in terms of the internal elastic ones $\mathbf{\tau}^{|e}$,
both operating in the current configuration and being numerically
coincident---cf. Eq. (\ref{sig=sigext+sigint})%
\begin{equation}
\mathbf{\tau}=\mathbf{\tau}^{|e}:\left.  \mathbb{M}_{d}^{d_{e}}\right\vert _{%
\lpo
}=\mathbf{\tau}^{|e}:\mathbb{I}^{S}=\mathbf{\tau}^{|e}
\label{tau=tauext+tauint}%
\end{equation}

Following analogous steps as in the small strain formulation, the dissipation
equation for the case when $\mathbf{l}_{p}\neq\mathbf{0}$, i.e. $\mathbf{d}%
_{e}=\,^{ct}\mathbf{d}_{e}$, becomes---compare to Eq.
(\ref{Dissipation epse ct})%
\begin{equation}
\mathcal{D}=-\mathbf{\tau}^{|e}:\,^{ct}\mathbf{d}_{e}>0\quad\text{if}%
\quad\mathbf{l}_{p}\neq\mathbf{0} \label{Dissipation de ct}%
\end{equation}
so we can define a flow rule in terms of an Eulerian plastic potential
$\phi_{\tau}$ through---compare to Eq. (\ref{flow rule small gradient ct})%
\begin{equation}
\,^{ct}\mathbf{d}_{e}=-\dot{\gamma}\frac{1}{k}\nabla\phi_{\tau}
\label{flow rule spatial gradient ct}%
\end{equation}
where $\dot{\gamma}$ is the plastic consistency parameter, $k$ the yield
stress and%
\begin{equation}
\nabla\phi_{\tau}:=\frac{\partial\phi_{\tau}\left(  \mathbf{\tau}%
^{|e};\mathbf{X}_{e}\right)  }{\partial\mathbf{\tau}^{|e}}
\label{stress gradient objective}%
\end{equation}
is the partial stress-gradient of the Eulerian potential $\phi_{\tau}$
performed with the spatial referential configuration of its arguments
remaining fixed, with $\phi_{\tau}\left(  \mathbf{\tau}^{|e};\mathbf{X}%
_{e}\right)  $ being an isotropic scalar-valued tensor function in its
arguments in the sense that $\phi_{\tau}(\mathbf{Q\tau}^{|e}\mathbf{Q}%
^{T};\mathbf{QX}_{e})=\phi_{\tau}\left(  \mathbf{\tau}^{|e};\mathbf{X}%
_{e}\right)  $, i.e. invariant under rigid body motions---cf. Ref.
\cite{MenzelSteinmann03} for an alternative, yet equivalent, interpretation.
Hence, exactly as in the small strain case, note that the associative flow
rule defined by Eq. (\ref{flow rule spatial gradient ct}) enforce a normal
projection onto the elastic domain in a continuum sense and that the plastic
spin does not explicitly take part in that six-dimensional equation, as one
would desire in a large strain context \cite{Lubliner86}. Clearly, the
internal elastic return is governed by the objective potential gradient
$\nabla\phi_{\tau}$ as given in Eq. (\ref{stress gradient objective}).

Positive dissipation is directly guaranteed in Eq. (\ref{Dissipation de ct})
if we choose---cf. Eq. (\ref{phi pequenas})%
\begin{equation}
\phi_{\tau}(\mathbf{\tau}^{|e};\mathbf{X}_{e})=\tfrac{1}{2}\mathbf{\tau}%
^{|e}:\mathbb{N}_{\tau}\left(  \mathbf{X}_{e}\right)  :\mathbf{\tau}^{|e}
\label{potential taue Xe}%
\end{equation}
with $\mathbb{N}_{\tau}=\mathbb{N}_{\tau}\left(  \mathbf{X}_{e}\right)  $
standing for an elastic-deformation-dependent symmetric positive-definite
fourth-order tensor lying in the same configuration as $\mathbf{d}_{e}$ and
$\mathbf{\tau}^{|e}$, i.e. the current configuration. For the reader
convenience, we refer to Eq. (\ref{Ntau NT}) below, where the tensor
$\mathbb{N}_{\tau}\left(  \mathbf{X}_{e}\right)  $ is explicitly defined in
terms of its Lagrangian-type logarithmic counterpart in the intermediate
configuration. Thus%
\begin{equation}
\nabla\phi_{\tau}=\frac{\partial\phi_{\tau}\left(  \mathbf{\tau}%
^{|e};\mathbf{X}_{e}\right)  }{\partial\mathbf{\tau}^{|e}}=\mathbb{N}_{\tau
}:\mathbf{\tau}^{|e}%
\end{equation}
and Eq. (\ref{Dissipation de ct}) reads ---cf. Eq. (\ref{Dissipationsige})%
\begin{equation}
\mathcal{D}=\frac{\mathbf{\tau}^{|e}:\mathbb{N}_{\tau}:\mathbf{\tau}^{|e}%
}{k^{2}}k\dot{\gamma}>0\quad\text{if}\quad\dot{\gamma}>0 \label{Diss eq}%
\end{equation}
The yield function $f_{\tau}(\mathbf{\tau}^{|e},k;\mathbf{X}_{e})$ and the
loading/unloading conditions are naturally identified in this last expression,
i.e. ---cf. Eq. (\ref{elastic response small})%
\begin{equation}
f_{\tau}(\mathbf{\tau}^{|e},k;\mathbf{X}_{e})=2\phi_{\tau}(\mathbf{\tau}%
^{|e};\mathbf{X}_{e})-k^{2}=\mathbf{\tau}^{|e}:\mathbb{N}_{\tau}:\mathbf{\tau
}^{|e}-k^{2}=0\quad\text{if}\quad\dot{\gamma}>0 \label{Yield fcn spatial}%
\end{equation}
and%
\begin{equation}
\dot{\gamma}=0\quad\text{if}\quad f_{\tau}(\mathbf{\tau}^{|e},k;\mathbf{X}%
_{e})=2\phi_{\tau}(\mathbf{\tau}^{|e};\mathbf{X}_{e})-k^{2}=\mathbf{\tau}%
^{|e}:\mathbb{N}_{\tau}:\mathbf{\tau}^{|e}-k^{2}<0
\end{equation}
whereupon we can write $\mathcal{D}=k\dot{\gamma}\geq0$ for $\dot{\gamma}%
\geq0$.

\begin{table}[h]
\caption{Finite strain multiplicative anisotropic elastoplasticity model.
Spatial description.}%
\label{Table - Spatial}
\begin{center}%
\begin{tabular}
[c]{|l|}\hline%
\begin{tabular}
[c]{l}%
\vspace{-0.1in}\\
(i) Multiplicative decomposition of the deformation gradient $\mathbf{X}%
=\mathbf{X}_{e}\mathbf{X}_{p}$\bigskip\\
(ii) Symmetric internal strain variable $\mathbf{a}_{e}(\mathbf{A}%
_{e};\mathbf{X}_{e})=\mathbf{X}_{e}^{-T}\mathbf{A}_{e}\mathbf{X}_{e}^{-1}%
$\bigskip\\
(iii) Kinematics induced by $\mathbf{X}_{e}(\mathbf{X},\mathbf{X}%
_{p})=\mathbf{XX}_{p}^{-1}$\medskip\\
\qquad$\overset{\diamond}{\mathbf{a}}_{e}=\mathbf{d}_{e}=\left.
\mathbf{d}_{e}\right\vert _{{\scriptsize \mathbf{l}_{p}=\mathbf{0}}}+\left.
\mathbf{d}_{e}\right\vert _{{\scriptsize \mathbf{d}=\mathbf{0}}}%
=\,^{tr}\mathbf{d}_{e}+\,^{ct}\mathbf{d}_{e}\neq\mathbf{d}-\mathbf{d}_{p}%
$\bigskip\\
(iv) Symmetric stresses deriving from the strain energy $\Psi_{A}%
(\mathbf{A}_{e})=\Psi_{a}(\mathbf{a}_{e};\mathbf{X}_{e})$\medskip\\
\qquad$\mathbf{\tau}^{|e}=\dfrac{\partial\Psi_{a}(\mathbf{a}_{e}%
;\mathbf{X}_{e})}{\partial\mathbf{a}_{e}}=\mathbf{X}_{e}\dfrac{d\Psi
_{A}(\mathbf{A}_{e})}{d\mathbf{A}_{e}}\mathbf{X}_{e}^{T}$\,,\qquad
$\mathbf{\tau}=\mathbf{\tau}^{|e}:\left.  \mathbb{M}_{d}^{d_{e}}\right\vert
_{{\scriptsize \mathbf{l}_{p}=\mathbf{0}}}\equiv\mathbf{\tau}^{|e}$\bigskip\\
(v) Evolution equation for associative symmetric plastic flow\medskip\\
\qquad$-\,^{ct}\mathbf{d}_{e}=\dot{\gamma}\dfrac{1}{k}\dfrac{\partial
\phi_{\tau}(\mathbf{\tau}^{|e};\mathbf{X}_{e})}{\partial\mathbf{\tau}^{|e}%
}\neq\mathbf{d}_{p}$\medskip\\
\qquad$\dot{\gamma}\geq0$\,,\quad$f_{\tau}(\mathbf{\tau}^{|e},k;\mathbf{X}%
_{e})=2\phi_{\tau}(\mathbf{\tau}^{|e};\mathbf{X}_{e})-k^{2}\leq0$\,,\quad
$\dot{\gamma}f_{\tau}(\mathbf{\tau}^{|e},k;\mathbf{X}_{e})=0$\bigskip\\
(vi) Additional evolution equation for skew-symmetric plastic flow
$\mathbf{w}_{p}$\medskip\\
{\small {Note: Potential $\Psi_{a}(\mathbf{a}_{e};\mathbf{X}_{e})$ and
function $f_{\tau}(\mathbf{\tau}^{|e},k;\mathbf{X}_{e})$ are anisotropic, in
general.}\bigskip}\\
\end{tabular}
\\\hline
\end{tabular}
\end{center}
\end{table}

\subsection{Dissipation inequality and flow rule in terms of spatial plastic
rates}

We can re-write Eq. (\ref{flow rule spatial gradient ct}) using Eq.
(\ref{de ct}) as%
\begin{equation}
sym\left(  \mathbf{X}_{e}\mathbf{l}_{p}\mathbf{X}_{e}^{-1}\right)
=\dot{\gamma}\frac{1}{k}\nabla\phi_{\tau}
\label{flow rule spatial gradient lp}%
\end{equation}
In the infinitesimal framework the internal variable being employed in the
evolution equation, whether elastic or plastic, is irrelevant in practice
---cf. Eqs. (\ref{flow rule small gradient}) and
(\ref{flow rule small plastic}). However in the finite strain case the
evolution of the internal variables, whether elastic or plastic, require very
different treatments, compare Eq. (\ref{flow rule spatial gradient ct}) with
Eq. (\ref{flow rule spatial gradient lp}).

We want also to remark that Eq. (\ref{flow rule spatial gradient lp}) is, in
essence, Eq. ($36.3$) of Ref. \cite{Simo98} (note that our $\mathbf{l}_{p}$ is
their $\mathbf{\bar{L}}_{p}$, see Eq. ($34.6$) in Ref. \cite{Simo98}), which
is further integrated therein with the plastic spin symmetrizing assumption
$skw\left(  \mathbf{X}_{e}\mathbf{l}_{p}\mathbf{X}_{e}^{-1}\right)
=\mathbf{0}$ by means of---cf. Table $36.1$ and Eqs. ($46.3$) and ($46.5$) in
Ref. \cite{Simo98}%
\begin{equation}
\mathbf{X}_{e}\mathbf{l}_{p}\mathbf{X}_{e}^{-1}=\dot{\gamma}\frac{1}{k}%
\nabla\phi_{\tau} \label{flow rule lp}%
\end{equation}
with $\mathbf{l}_{p}=\mathbf{\dot{X}}_{p}\mathbf{X}_{p}^{-1}$ using our
notation, in order to arrive at an algorithmic formulation based on internal
elastic variables upon considering an exponential mapping approximation, cf.
Eq. ($46.9$a) in Ref. \cite{Simo98}. Indeed, Eq. ($46.3$) of Ref.
\cite{Simo98} (Eq. (\ref{flow rule lp})) is interpreted therein to be written
in \textquotedblleft non-standard form\textquotedblright\ due to the fact that
\textquotedblleft the time derivative is hidden in the definition of the
spatial plastic rate\textquotedblright\ \cite{Simo98}, i.e. $\mathbf{l}%
_{p}=\mathbf{\dot{X}}_{p}\mathbf{X}_{p}^{-1}$ using our notation. On the
contrary, we herein interpret Eq. (\ref{flow rule spatial gradient ct}) to be
written in \emph{standard form} if one considers corrector elastic rates
(whether infinitesimal, Eulerian or Lagrangian) rather than plastic rates,
recall the interpretation given above in Eq.
(\ref{flow rule small gradient ct}) within the small strain setting and
see\ below the description in the intermediate configuration. The reader can
compare again Eqs. (\ref{flow rule small gradient}) and
(\ref{flow rule small plastic}) and, in the light of the above lines see that
they both indeed present clearly different views of the physics behind the
same problem. This observation is again parallel to that presented in large
strain viscoelasticity \cite{LatMonCM2015}\ where the use of the novel
approach allowed for the development of phenomenological anisotropic
formulations valid for large deviations from thermodynamic equilibrium.

\subsection{Comparison with other formulations which are restricted to
isotropy}

In isotropic finite strain elastoplasticity formulations it is frequent the
case in which the internal evolution equations in spatial description are
expressed in terms of the Lie derivative of the elastic left
Cauchy--Green-like deformation tensor \cite{PericDettmer03,BonetWoodBook}, an
approach that goes back to the works of Sim\'{o} and Miehe
\cite{SimoMiehe92,Simo92}. An analogous setting is encountered in isotropic
finite strain viscoelasticity and viscoplasticity formulations
\cite{ReeseGovindjee98,PericDettmer03,HolmesLougram10}. We take advantage
herein of the previous concepts of partial differentiation and mapping tensors
in order to interpret some terms involving the Lie derivative operator. The
left Cauchy--Green-like tensor $\mathbf{B}_{e}=\mathbf{X}_{e}\mathbf{X}%
_{e}^{T}$ may be considered a function of the deformation gradient tensor
$\mathbf{X}$ and the inverse of the plastic right Cauchy--Green deformation
tensor $\mathbf{C}_{p}^{-1}=\mathbf{X}_{p}^{-1}\mathbf{X}_{p}^{-T}$ as---we
separate the arguments by a comma because $\mathbf{X}$ and $\mathbf{C}%
_{p}^{-1}$ represent two different deformation processes, cf. Eq.
(\ref{epse dependence 0})%

\begin{equation}
\mathbf{B}_{e}(\mathbf{X},\mathbf{C}_{p}^{-1})=\mathbf{XC}_{p}^{-1}%
\mathbf{X}^{T}=\mathbf{X}\odot\mathbf{X}:\mathbf{C}_{p}^{-1}%
\end{equation}
The partial contribution to the total rate of $\mathbf{B}_{e}$ when
$\mathbf{X}$ is frozen stands for the Lie derivative of $\mathbf{B}_{e}$
relative to the total deformation field \cite{LatMonCM2015}%
\begin{equation}
\left.  \mathbf{\dot{B}}_{e}\right\vert _{%
\xdo
}=\left.  \frac{\partial\mathbf{B}_{e}}{\partial\mathbf{C}_{p}^{-1}%
}\right\vert _{%
\xdo
}:\mathbf{\dot{C}}_{p}^{-1}=\mathbf{X}\odot\mathbf{X}:\mathbf{\dot{C}}%
_{p}^{-1}=\mathbf{X\dot{C}}_{p}^{-1}\mathbf{X}^{T}=\mathcal{L}\mathbf{B}_{e}%
\end{equation}
where $\mathbf{\dot{C}}_{p}^{-1}:=d\mathbf{C}_{p}^{-1}/dt$. We also have%
\begin{equation}
\tfrac{1}{2}\mathcal{L}\mathbf{B}_{e}=\tfrac{1}{2}\mathbf{X\dot{C}}_{p}%
^{-1}\mathbf{X}^{T}=-\mathbf{X}_{e}\mathbf{\mathbf{d}}_{p}\mathbf{X}_{e}^{T}%
\end{equation}

Consider now the functional dependence $\mathbf{l}_{e}\left(  \mathbf{l}%
,\mathbf{l}_{p}\right)  =\mathbf{l}-\mathbf{X}_{e}\mathbf{l}_{p}\mathbf{X}%
_{e}^{-1}$ obtained from Eq. (\ref{l=le+lp}). If we additionally assume that
the plastic spin in the intermediate configuration $\mathbf{w}_{p}=skw\left(
\mathbf{l}_{p}\right)  $ vanishes, we arrive at%
\begin{equation}
\left.  \mathbf{l}_{e}\right\vert _{%
\lowpo
}=-\mathbf{X}_{e}\mathbf{d}_{p}\mathbf{X}_{e}^{-1}=\tfrac{1}{2}(\mathcal{L}%
\mathbf{B}_{e})\mathbf{B}_{e}^{-1}%
\end{equation}
so we may interpret the term $\tfrac{1}{2}(\mathcal{L}\mathbf{B}%
_{e})\mathbf{B}_{e}^{-1}$ as the partial (corrector) contribution to the
elastic velocity gradient $\mathbf{l}_{e}$ when both $\mathbf{l}=\mathbf{0}$
and $\mathbf{w}_{p}=\mathbf{0}$. Indeed, this last equation is the
generalization of, for example, Eq. ($7.18$) of Ref. \cite{BonetWoodBook},
where the simplifying hypothesis of isotropy is previously made to arrive at
that result, see Eqs. ($7.7$) of the same Reference.

The dissipation inequality given in Eq. (\ref{Dissipation de ct}) reads%

\begin{equation}
\mathcal{D}=-\mathbf{\tau}^{|e}:\left.  \mathbf{d}_{e}\right\vert _{%
\doo
}=-\mathbf{\tau}:\left.  \mathbf{l}_{e}\right\vert _{%
\lo
}>0\quad\text{if}\quad\mathbf{l}_{p}\neq\mathbf{0}%
\end{equation}
where we have used the fact that $\mathbf{\tau}^{|e}=\mathbf{\tau}$ is
symmetric. If we additionally prescribe a vanishing plastic spin, i.e.
$\mathbf{w}_{p}=\mathbf{0}$, the dissipation inequality reads%
\begin{equation}
\mathcal{D}=-\mathbf{\tau}:\left.  \mathbf{l}_{e}\right\vert _{%
\lowpo
}=-\mathbf{\tau}:\tfrac{1}{2}(\mathcal{L}\mathbf{B}_{e})\mathbf{B}_{e}%
^{-1}>0\quad\text{if}\quad\mathbf{d}_{p}\neq\mathbf{0}%
\end{equation}
which, note, is still valid for anisotropic elastoplasticity. A possible flow
rule is%
\begin{equation}
-sym\left(  \tfrac{1}{2}(\mathcal{L}\mathbf{B}_{e})\mathbf{B}_{e}^{-1}\right)
=-sym(\left.  \mathbf{l}_{e}\right\vert _{%
\lowpo
})=-\left.  \mathbf{d}_{e}\right\vert _{%
\dowpo
}=\dot{\gamma}\frac{1}{k}\mathbb{\nabla}\phi_{\tau}%
\end{equation}
which is the general flow rule of Eq. (\ref{flow rule spatial gradient ct})
when we add the simplifying assumption $\mathbf{w}_{p}=\mathbf{0}$. We remark
that we have arrived at the same evolution equation in terms of $\mathbf{d}%
_{e}$ considering either $\mathbf{w}_{p}\neq\mathbf{0}$ or $\mathbf{w}%
_{p}=\mathbf{0}$, which means that the return to the elastic domain is,
effectively, independent of the plastic spin $\mathbf{w}_{p}$ in the
intermediate configuration. An additional, independent constitutive equation
for $\mathbf{w}_{p}$ would be needed in order to describe the simultaneous
evolution of the intermediate configuration.

Finally, if the simplifying assumption of isotropic elasticity is made,
$\mathbf{B}_{e}$ commutes with $\mathbf{\tau}=\mathbf{\tau}^{|e}%
=2(d\Psi\left(  \mathbf{B}_{e}\right)  /d\mathbf{B}_{e})\mathbf{B}_{e}$. If we
additionally assume isotropic plastic behavior, then $\mathbf{B}_{e}$ also
commutes with both $\mathbb{\nabla}\phi_{\tau}=\mathbb{N}_{\tau}:\mathbf{\tau
}^{|e}$ and $\mathcal{L}\mathbf{B}_{e}$ and we recover the well-known,
although \textquotedblleft non-conventional\textquotedblright\ (recall remark
in \cite{Simo92}), local evolution equation for $\mathbf{B}_{e}$
\cite{SimoMiehe92}%
\begin{equation}
-\tfrac{1}{2}\mathcal{L}\mathbf{B}_{e}=\dot{\gamma}\frac{1}{k}(\mathbb{\nabla
}\phi_{\tau})\mathbf{B}_{e} \label{flow rule lie}%
\end{equation}
which can be integrated in principal spatial directions, as originally, or
applying a much more efficient integration procedure in the case of the
neo-Hookean strain energy function \cite{ShuLanIhl13}. The reader can now
compare the simplicity of the interpretation of Eq.
(\ref{flow rule spatial gradient ct}) of general validity with the arguably
more elusive one of Eq. (\ref{flow rule lie}), which is furthermore restricted
to isotropy.

\section{Finite strain anisotropic elastoplasticity formulated in the
intermediate configuration: the common approach in the literature}

As aforementioned, in the finite strain case the description of the internal
variables evolution, whether elastic or plastic, require very different
treatments, recall Eqs. (\ref{flow rule spatial gradient ct}) and
(\ref{flow rule spatial gradient lp}) in the spatial description. Models for
anisotropic multiplicative elastoplasticity are commonly formulated in the
intermediate configuration using evolution equations for internal variables
that are plastic in nature, typically the plastic deformation gradient
$\mathbf{X}_{p}$. We briefly discuss this approach in this section.

\subsection{Dissipation inequality and flow rule in terms of $\mathbf{l}_{p}$}

Consider Eq. (\ref{Dissipation de ct}) written in terms of the plastic
velocity gradient $\mathbf{l}_{p}$ rather than in terms of the corrector-type
elastic deformation rate tensor $\mathbf{d}_{e}|_{%
\doo
}=\,^{ct}\mathbf{d}_{e}$%
\begin{equation}
\mathcal{D}=-\mathbf{\tau}^{|e}:\left.  \mathbb{M}_{l_{p}}^{d_{e}}\right\vert
_{%
\doo
}:\mathbf{l}_{p}>0\quad\text{if}\quad\mathbf{l}_{p}\neq\mathbf{0}%
\end{equation}
where $\left.  \mathbb{M}_{l_{p}}^{d_{e}}\right\vert _{%
\doo
}$ is the mapping tensor already defined in Eq. (\ref{de ct}). We can define
the power-conjugate stress tensor of $\mathbf{l}_{p}$ as%
\begin{equation}
\mathbf{\Xi}^{|e}:=-\mathbf{\tau}^{|e}:\left.  \mathbb{M}_{l_{p}}^{d_{e}%
}\right\vert _{%
\doo
}=\frac{1}{2}\mathbf{\tau}^{|e}:\left(  \mathbf{X}_{e}\odot\mathbf{X}_{e}%
^{-T}+\mathbf{X}_{e}^{-T}\boxdot\mathbf{X}_{e}\right)  =\mathbf{X}_{e}%
^{T}\mathbf{\tau}^{|e}\mathbf{X}_{e}^{-T} \label{mandel tau}%
\end{equation}
and using $\mathbf{\tau}^{|e}=\mathbf{X}_{e}\mathbf{S}^{|e}\mathbf{X}_{e}^{T}$%
\begin{equation}
\mathbf{\Xi}^{|e}=\mathbf{C}_{e}\mathbf{S}^{|e} \label{mandel S}%
\end{equation}
which is the common definition of the non-symmetric Mandel stress tensor in
the intermediate configuration. The dissipation inequality is then%
\begin{equation}
\mathcal{D}=\mathbf{\Xi}^{|e}:\mathbf{l}_{p}>0\quad\text{if}\quad
\mathbf{l}_{p}\neq\mathbf{0} \label{Dissipation mandel}%
\end{equation}
which is fulfilled automatically employing the following nine-dimensional flow
rule---originally proposed by Mandel \cite{Mandel74}%
\begin{equation}
\mathbf{l}_{p}=\dot{\gamma}\frac{1}{k}\nabla\phi_{\Xi}
\label{flow rule mandel lp}%
\end{equation}
with%
\begin{equation}
\phi_{\Xi}=\tfrac{1}{2}\mathbf{\Xi}^{|e}:\mathbb{N}_{\Xi}:\mathbf{\Xi}^{|e}%
\end{equation}
where $\mathbb{N}_{\Xi}$ is a positive-definite tensor with major symmetries
but lacking minor symmetries. The added difficulty associated to the
integration of this type of non-symmetric evolution equations for the plastic
velocity gradient $\mathbf{l}_{p}$ is apparent \cite{SansourKarsajSoric06}.
The experimental determination of the yield parameters included in
$\mathbb{N}_{\Xi}$ implies the consideration of additional tests with respect
to the case in which a six-dimensional flow rule is considered. Furthermore,
note that the plastic spin $\mathbf{w}_{p}=skw\left(  \mathbf{l}_{p}\right)  $
is given from $skw(\nabla\phi_{\Xi})$ in Eq. (\ref{flow rule mandel lp}) as an
additional assumption \cite{Lubliner86}, which is a crucial difference with
the small strain formulation.

\subsection{Dissipation inequality and flow rule in terms of $\mathbf{l}_{p}$
with $\mathbf{w}_{p}=\mathbf{0}$}

Plastic spin effects can be important in finite strain anisotropic plasticity
\cite{MontansBathe07}. However, the constitutive equation for the plastic spin
$\mathbf{w}_{p}=\mathbf{0}$ is frequently considered in Eq.
(\ref{Dissipation mandel}). This simplifying assumption leads to the following
dissipation inequality---we define $\mathbf{\Xi}_{s}^{|e}=sym(\mathbf{\Xi
}^{|e})$%
\begin{equation}
\mathcal{D}=\mathbf{\Xi}^{|e}:\mathbf{d}_{p}=\mathbf{\Xi}_{s}^{|e}%
:\mathbf{d}_{p}>0\quad\text{if}\quad\mathbf{d}_{p}\neq\mathbf{0}%
\end{equation}
and to the following six-dimensional anisotropic flow rule for the plastic
deformation rate tensor---see \cite{EidelGruttman03}%
\cite{CamineroMontansBathe11} among many others%
\begin{equation}
\mathbf{l}_{p}=\mathbf{\dot{X}}_{p}\mathbf{X}_{p}^{-1}=\mathbf{d}_{p}%
=\dot{\gamma}\frac{1}{k}\nabla\phi_{\Xi_{s}} \label{flow rule mandelsym dp}%
\end{equation}
In the present context, one can now take%
\begin{equation}
\phi_{\Xi_{s}}=\tfrac{1}{2}\mathbf{\Xi}_{s}^{|e}:\mathbb{N}_{\Xi_{s}%
}:\mathbf{\Xi}_{s}^{|e}%
\end{equation}
with $\mathbb{N}_{\Xi_{s}}$ being fully symmetric and positive definite, so%
\begin{equation}
\mathcal{D}=\dot{\gamma}\frac{1}{k}\mathbf{\Xi}_{s}^{|e}:\mathbb{N}_{\Xi_{s}%
}:\mathbf{\Xi}_{s}^{|e}\geq0\quad\text{for}\quad\dot{\gamma}\geq0
\end{equation}
which, following already customary steps, naturally defines the yield function
$f_{\Xi_{s}}(\mathbf{\Xi}_{s}^{|e},k)=\mathbf{\Xi}_{s}^{|e}:\mathbb{N}%
_{\Xi_{s}}:\mathbf{\Xi}_{s}^{|e}-k^{2}=0$ for $\dot{\gamma}>0$.

If the hyperelastic response is modelled with the Hencky strain energy
function in the intermediate configuration and the additional restriction to
moderately large elastic deformations is taken, then $\mathbf{\Xi}_{s}^{|e}$
is, in practice, the work-conjugate stress tensor of the elastic logarithmic
strains in the intermediate configuration $\mathbf{E}_{e}=\frac{1}{2}%
\ln(\mathbf{X}_{e}^{T}\mathbf{X}_{e})$ \cite{CamineroMontansBathe11}. This
consideration greatly facilitates the algorithmic implementation of this
formulation based on the evolution of the plastic gradient tensor
$\mathbf{X}_{p}$ by means of Eq. (\ref{flow rule mandelsym dp}), retaining at
the same time the main features of the isotropic logarithmic-strain-based
formulation of Ref. \cite{EterovicBathe90}.

Consider now the isotropic elasticity case, for which elastic strains and
stresses commute. Then, the Mandel stress tensor, as given in Eq.
(\ref{mandel tau}), simplifies to the internal, elastically rotated Kirchhoff
stress tensor---we introduce herein the left polar decomposition of the
elastic deformation gradient $\mathbf{X}_{e}=\mathbf{V}_{e}\mathbf{R}_{e}$%
\begin{equation}
\mathbf{\Xi}^{|e}=\mathbf{X}_{e}^{T}\mathbf{\tau}^{|e}\mathbf{X}_{e}%
^{-T}=\mathbf{R}_{e}^{T}\mathbf{V}_{e}\mathbf{\tau}^{|e}\mathbf{V}_{e}%
^{-1}\mathbf{R}_{e}=\mathbf{R}_{e}^{T}\mathbf{\tau}^{|e}\mathbf{R}%
_{e}=:\mathbf{\tau}_{R}^{|e} \label{rotated kirchhoff stresses}%
\end{equation}
which is a symmetric tensor. Then we can rephrase the potential $\phi_{\Xi}$
as%
\begin{equation}
\phi_{\Xi}\equiv\phi_{\tau}=\tfrac{1}{2}\mathbf{\tau}_{R}^{|e}:\mathbb{N}%
_{\tau}^{R}:\mathbf{\tau}_{R}^{|e}%
\end{equation}
with $\mathbb{N}_{\tau}^{R}$ being fully symmetric, but not necessarily
isotropic. Thus---note that this equation implies $\mathbf{w}_{p}=\mathbf{0}$%
\begin{equation}
\mathbf{\dot{X}}_{p}=\dot{\gamma}\frac{1}{k}(\nabla\phi_{\tau})\mathbf{X}_{p}%
\end{equation}
which is, in essence, the flow rule (originally proposed for isotropic
plasticity) of Weber and Anand \cite{WeberAnand90}\ and Eterovic and Bathe
\cite{EterovicBathe90}. However, note that it can also be used with
anisotropic plasticity \cite{MontansBathe05}.

\section{Finite strain anisotropic elastoplasticity formulated in the
intermediate configuration: our different proposed approach}

We present in this section a new framework for finite strain anisotropic
elastoplasticity formulated in the intermediate configuration in which the
basic internal variables are Lagrangian-like elastic measures consistent with
the multiplicative decomposition. We show that similar functional dependencies
to those used within the small strain theory may be established. The concepts
of partial differentiation, mapping tensors and the trial-corrector elastic
decomposition are firstly applied, just for motivation, to quadratic strains
due to its analytical simplicity. An equivalent analysis in terms of
logarithmic strain measures will allow us to derive a fully Lagrangian
elastoplastic formulation in the intermediate configuration with an apparent
similarity to the small strain one.

\subsection{Kinematic description in terms of $\,^{ct}\mathbf{\dot{A}}_{e}$}

From the Lee decomposition of Eq. (\ref{LeeDecomposition}), the total
Green--Lagrange strains in the reference configuration and the elastic
Green--Lagrange-like strains in the intermediate configuration are
$\mathbf{A}:=\frac{1}{2}(\mathbf{X}^{T}\mathbf{X}-\mathbf{I})$ and
$\mathbf{A}_{e}:=\frac{1}{2}(\mathbf{X}_{e}^{T}\mathbf{X}_{e}-\mathbf{I})$.
Following the idea introduced for small strains, and further applied to
spatial deformation rate tensors, we write the dependent, internal elastic
variable $\mathbf{A}_{e}$ as a function of the independent, external variable
$\mathbf{A}$ and the independent, internal plastic variable $\mathbf{X}_{p}$
as%
\begin{equation}
\mathbf{A}_{e}\left(  \mathbf{A},\mathbf{X}_{p}\right)  =\mathbf{X}_{p}%
^{-T}\left(  \mathbf{A}-\mathbf{A}_{p}\right)  \mathbf{X}_{p}^{-1}%
=\mathbf{X}_{p}^{-T}\odot\mathbf{X}_{p}^{-T}:\left(  \mathbf{A}-\mathbf{A}%
_{p}\right)  \label{Ae(A,Xp)}%
\end{equation}
where the plastic Green--Lagrange strain tensor is defined in the reference
configuration as $\mathbf{A}_{p}:=\frac{1}{2}(\mathbf{X}_{p}^{T}\mathbf{X}%
_{p}-\mathbf{I})$. The total rate of $\mathbf{A}_{e}$ may be written applying
the chain rule of differentiation to the tensor-valued function of two
tensor-valued variables $\mathbf{A}_{e}\left(  \mathbf{A},\mathbf{X}%
_{p}\right)  $ as%
\begin{equation}
\mathbf{\dot{A}}_{e}=\left.  \frac{\partial\mathbf{A}_{e}}{\partial\mathbf{A}%
}\right\vert _{%
\xpdo
}:\mathbf{\dot{A}}+\left.  \frac{\partial\mathbf{A}_{e}}{\partial
\mathbf{X}_{p}}\right\vert _{%
\ado
}:\mathbf{\dot{X}}_{p} \label{Aedot}%
\end{equation}
where identifying terms, and for further use, we obtain the fourth-order
\emph{partial} gradient\ tensor---compare to the identity mapping tensor
present in Eq. (\ref{de tr})%
\begin{equation}
\frac{\partial\mathbf{A}_{e}\left(  \mathbf{A},\mathbf{X}_{p}\right)
}{\partial\mathbf{A}}\equiv\left.  \frac{\partial\mathbf{A}_{e}}%
{\partial\mathbf{A}}\right\vert _{%
\xpdo
}=\mathbf{X}_{p}^{-T}\odot\mathbf{X}_{p}^{-T}\equiv\left.  \mathbb{M}_{\dot
{A}}^{\dot{A}_{e}}\right\vert _{%
\xpdo
} \label{dAe/dA _Xpdot=0}%
\end{equation}
The fourth-order tensor of Eq. (\ref{dAe/dA _Xpdot=0}) is a purely geometrical
tensor in the sense that it is known at any given deformation state in which
the Lee factorization is known. The total rate of $\mathbf{A}_{e}$ in Eq.
(\ref{Aedot}) may also be interpreted as the addition of the two independent
\emph{trial} and \emph{corrector} contributions%
\begin{equation}
\mathbf{\dot{A}}_{e}=\left.  \mathbf{\dot{A}}_{e}\right\vert _{%
\xpdo
}+\left.  \mathbf{\dot{A}}_{e}\right\vert _{%
\ado
}=\,^{tr}\mathbf{\dot{A}}_{e}+\,^{ct}\mathbf{\dot{A}}_{e} \label{AedotSplit}%
\end{equation}
Hence, and for further comparison with the logarithmic-based formulation, note
that the fourth-order tensor of Eq. (\ref{dAe/dA _Xpdot=0}) furnishes the
proper push-forward mapping over $\mathbf{\dot{A}}$, lying in the reference
configuration, to give $\,^{tr}\mathbf{\dot{A}}_{e}$ (i.e. $\mathbf{\dot{A}%
}_{e}$ with $\mathbf{\dot{X}}_{p}=\mathbf{0}$), lying in the intermediate
configuration. Importantly, Equations (\ref{Aedot}) and (\ref{AedotSplit}) are
fully consistent with the multiplicative decomposition of the deformation
gradient, whereas the add-hoc plastic metric decomposition%
\begin{equation}
\mathbf{\dot{A}}_{e}=\mathbf{\dot{A}}-\mathbf{\dot{A}}_{p}%
\end{equation}
is not consistent with multiplicative plasticity, in general, recall Eq.
(\ref{Ae(A,Xp)}).

\subsection{Dissipation inequality and flow rule in terms of natural corrector
elastic strain rates}

We now draw our attention to the arguably more natural logarithmic strain
framework, which we favour because of the natural properties of those strain
measures \cite{Anand79}, \cite{Anand86}, \cite{Rice75}, \cite{RolphBathe84},
\cite{LatMonIJSS2014}, \cite{NeffGhiba16}. At large strains, both quadratic
and Hencky strains are related by one-to-one mapping tensors
\cite{LatMonAPM2016}. Consider the \emph{explicit} analytical dependence
$\mathbf{A}_{e}\left(  \mathbf{A},\mathbf{X}_{p}\right)  $ given in Eq.
(\ref{Ae(A,Xp)}). Since the one-to-one, purely kinematical relations
$\mathbf{A}_{e}=\mathbf{A}_{e}\left(  \mathbf{E}_{e}\right)  $ and
$\mathbf{A}=\mathbf{A}\left(  \mathbf{E}\right)  $ hold, where $\mathbf{E}%
_{e}=\frac{1}{2}\ln(\mathbf{X}_{e}^{T}\mathbf{X}_{e})$ and $\mathbf{E}%
=\frac{1}{2}\ln(\mathbf{X}^{T}\mathbf{X})$ are the elastic and total material
logarithmic strain tensors in their respective configurations, we have also
the generally \emph{implicit} dependence $\mathbf{E}_{e}\left(  \mathbf{E}%
,\mathbf{X}_{p}\right)  $. Hence, analogously to Eq. (\ref{Aedot}), we can
decompose the internal elastic logarithmic strain rate tensor $\mathbf{\dot
{E}}_{e}$ by means of the addition of two partial contributions---cf. Eq.
(\ref{epsedot})%
\begin{equation}
\mathbf{\dot{E}}_{e}=\left.  \frac{\partial\mathbf{E}_{e}}{\partial\mathbf{E}%
}\right\vert _{%
\xpdo
}:\mathbf{\dot{E}}+\left.  \frac{\partial\mathbf{E}_{e}}{\partial
\mathbf{X}_{p}}\right\vert _{%
\edo
}:\mathbf{\dot{X}}_{p}=\left.  \mathbf{\dot{E}}_{e}\right\vert _{%
\xpdo
}+\left.  \mathbf{\dot{E}}_{e}\right\vert _{%
\edo
} \label{Eedot}%
\end{equation}
As in the small strain case, this decomposition in rate form is the origin of
the operator split typically employed for elastic internal variables in
computational inelasticity within an algorithmic framework. As well known,
this operator split consists of a trial elastic predictor, for which
$\mathbf{X}_{p}$ is frozen, and a plastic corrector, for which $\mathbf{E}$ is
frozen. The reader is again referred to Ref. \cite{LatMonCM2015} for an
algorithmic implementation of this type in the context of viscoelasticity.
Accordingly, we define the \emph{trial} and \emph{corrector} contributions to
$\mathbf{\dot{E}}_{e}$ within the finite strain continuum theory as---cf. Eqs.
(\ref{epsedot tr ct})%
\begin{equation}
\mathbf{\dot{E}}_{e}=\left.  \mathbf{\dot{E}}_{e}\right\vert _{%
\xpdo
}+\left.  \mathbf{\dot{E}}_{e}\right\vert _{%
\edo
}=:\,^{tr}\mathbf{\dot{E}}_{e}+\,^{ct}\mathbf{\dot{E}}_{e} \label{Eedot tr ct}%
\end{equation}
i.e., for a given state of deformation $\mathbf{X}=\mathbf{X}_{e}%
\mathbf{X}_{p}$ at a given instant, the trial elastic contribution
$\,^{tr}\mathbf{\dot{E}}_{e}$ to the total elastic logarithmic strain rate
$\mathbf{\dot{E}}_{e}$ depends on the total logarithmic strain rate
$\mathbf{\dot{E}}$ only (i.e. $\mathbf{X}_{p}$ is frozen) and the plastic
corrector contribution $\,^{ct}\mathbf{\dot{E}}_{e}$ to the total elastic
logarithmic strain rate $\mathbf{\dot{E}}_{e}$ depends on the total plastic
deformation gradient rate $\mathbf{\dot{X}}_{p}$ only (i.e. $\mathbf{E}$ is frozen).

We want to remark that the general expression in rate form given in Eq.
(\ref{Eedot})$_{1}$ particularizes to%
\begin{equation}
\mathbf{\dot{E}}_{e}=\mathbf{\dot{E}}-\mathbf{\dot{E}}_{p}
\label{plasticMetric}%
\end{equation}
in very few special cases only, e.g. axial loadings along preferred axes in
orthotropic materials. Hence, formulations based on ad-hoc decompositions of
the form $\mathbf{E}_{e}=\mathbf{E}-\mathbf{E}_{p}$ involving the so-called
plastic metric (from which Eq. (\ref{plasticMetric}) is immediately derived),
cf. \cite{PapadopoulusLu98}, \cite{Miehe02}, \cite{SansourWagner03},
\cite{Ulz09} and also \cite{Schmidt05}, are not generally consistent with the
continuum kinematic formulation derived from the Lee decomposition which is
represented by Eq. (\ref{Eedot}) in the most general case and that we use in
the present work, and analogously in Ref. \cite{LatMonCM2015}, without further simplifications.

The dissipation inequality written in terms of Lagrangian logarithmic strains
can be seemingly obtained from Eq. (\ref{Dissipation spatial general}) as%
\begin{equation}
\mathcal{D}=\mathcal{P}-\dot{\Psi}_{E}=\mathbf{T}:\mathbf{\dot{E}}%
-\mathbf{T}^{|e}:\mathbf{\dot{E}}_{e}\geq0
\end{equation}
where $\Psi_{E}\left(  \mathbf{E}_{e}\right)  $ is the orthotropic strain
energy function given in this case in terms of elastic logarithmic
strains---with the simplifying assumption $\mathbf{\dot{a}}_{1}=\mathbf{\dot
{a}}_{2}=\mathbf{\dot{a}}_{3}=\mathbf{0}$%
\begin{equation}
\Psi_{E}=\Psi_{E}\left(  \mathbf{E}_{e},\mathbf{a}_{1}\otimes\mathbf{a}%
_{1},\mathbf{a}_{2}\otimes\mathbf{a}_{2}\right)
\end{equation}
and%
\begin{equation}
\mathbf{T}^{|e}=\frac{d\Psi_{E}\left(  \mathbf{E}_{e}\right)  }{d\mathbf{E}%
_{e}} \label{Tinte}%
\end{equation}
is the internal generalized Kirchhoff stress tensor that directly derives from
$\Psi_{E}\left(  \mathbf{E}_{e}\right)  $, which is the work-conjugate stress
tensor of $\mathbf{E}_{e}$ in the most general case \cite{LatMonAPM2016}.

Following the already customary arguments, if $\mathbf{\dot{X}}_{p}%
=\mathbf{0}$ we have $\mathbf{\dot{E}}_{e}\equiv\mathbf{\dot{E}}_{e}|_{%
\xpdo
}=\,^{tr}\mathbf{\dot{E}}_{e}$ and%
\begin{equation}
\mathcal{D}=\mathbf{T}:\mathbf{\dot{E}}-\mathbf{T}^{|e}:\,^{tr}\mathbf{\dot
{E}}_{e}=0\quad\text{if}\quad\mathbf{\dot{X}}_{p}=\mathbf{0}%
\end{equation}
so we arrive at---cf. Eq. (\ref{sig=sigext+sigint})%
\begin{equation}
\mathbf{T}=\mathbf{T}^{|e}:\left.  \frac{\partial\mathbf{E}_{e}}%
{\partial\mathbf{E}}\right\vert _{%
\xpdo
}=\left.  \frac{\partial\Psi_{E}\left(  \mathbf{E}_{e}\right)  }%
{\partial\mathbf{E}}\right\vert _{%
\xpdo
} \label{TTextTint}%
\end{equation}
with the fourth-order tensor $\partial\mathbf{E}_{e}/\partial\mathbf{E|}_{%
\xpdo
}$, present in Eq. (\ref{Eedot}), furnishing the proper mappings between
$\mathbf{\dot{E}}$ and $\,^{tr}\mathbf{\dot{E}}_{e}$ and also between
$\mathbf{T}^{|e}$ and $\mathbf{T}$ when the intermediate configuration remains
fixed, so%
\begin{equation}
\left.  \dot{\Psi}_{E}\right\vert _{%
\xpdo
}=\,^{tr}\dot{\Psi}_{E}=\mathbf{T}^{|e}:\,^{tr}\mathbf{\dot{E}}_{e}%
=\mathbf{T}^{|e}:\left.  \frac{\partial\mathbf{E}_{e}}{\partial\mathbf{E}%
}\right\vert _{%
\xpdo
}:\mathbf{\dot{E}}=\mathbf{T}:\mathbf{\dot{E}}%
\end{equation}

On the other side, the dissipation equation whenever $\mathbf{\dot{X}}_{p}%
\neq\mathbf{0}$ reduces to---cf. Eq. (\ref{Dissipation epse ct})%
\begin{equation}
\mathcal{D}=-\mathbf{T}^{|e}:\,^{ct}\mathbf{\dot{E}}_{e}>0\quad\text{if}%
\quad\dot{\gamma}>0 \label{DissipationEe}%
\end{equation}
The following flow rule may be chosen---cf. Eq.
(\ref{flow rule small gradient ct})%
\begin{equation}
\,^{ct}\mathbf{\dot{E}}_{e}=-\dot{\gamma}\frac{1}{k}\mathbb{\nabla}\phi_{T}
\label{flow rule logarithmic ct}%
\end{equation}
where $\phi_{T}(\mathbf{T}^{|e})$ is a Lagrangian internal potential function.
The convex potential%
\begin{equation}
\phi_{T}(\mathbf{T}^{|e})=\tfrac{1}{2}\mathbf{T}^{|e}:\mathbb{N}%
_{T}:\mathbf{T}^{|e}%
\end{equation}
automatically fulfills the physical requirement%
\begin{equation}
\mathcal{D}=\dot{\gamma}\frac{1}{k}\mathbf{T}^{|e}:\mathbb{N}_{T}%
:\mathbf{T}^{|e}>0\quad\text{if}\quad\dot{\gamma}>0 \label{Dissipation Te}%
\end{equation}
when $\mathbb{N}_{T}$ is a positive-definite fully symmetric fourth order
tensor. Note that Eq. (\ref{flow rule logarithmic ct}) provokes the
instantaneous closest-point projection to the elastic domain in a continuum
sense in the logarithmic space. Furthermore, consistently with the normality
rule emanating from the principle of maximum dissipation \cite{Lubliner86},
the plastic spin in the intermediate configuration $\mathbf{w}_{p}$ does not
take \emph{explicit} part in Eq. (\ref{flow rule logarithmic ct}). Once the
hyperelastic stress-strain relations are assumed and a yield condition is
postulated, the associative flow rule given in Eq.
(\ref{flow rule logarithmic ct}) can be integrated independently of the
plastic spin evolution. In this respect, note that the direct integration of
Eq. (\ref{flow rule logarithmic ct}) in terms of the symmetric internal
elastic strain variable $\mathbf{E}_{e}$ during the corresponding algorithmic
corrector phase is completely equivalent to the (certainly more challenging)
integration of the following evolution equation for $\mathbf{\dot{X}}%
_{p}=\mathbf{l}_{p}\mathbf{X}_{p}$ ---see second addends in Eq. (\ref{Eedot})%
\begin{equation}
\left.  \frac{\partial\mathbf{E}_{e}}{\partial\mathbf{X}_{p}}\right\vert _{%
\edo
}:(\mathbf{d}_{p}+\mathbf{w}_{p})\mathbf{X}_{p}=-\dot{\gamma}\frac{1}%
{k}\mathbb{\nabla}\phi_{T}%
\end{equation}
Once the symmetric flow given by Eq. (\ref{flow rule logarithmic ct}) is
integrated, the intermediate configuration, defined by $\mathbf{X}_{p}$,
remains undetermined up to an arbitrary finite rotation $\mathbf{R}_{e}$
\cite{Simo88}, which may be finally updated during the convergence phase for
the computation of the next incremental load step, as we already did in a
similar multiplicative framework based on the Sidoroff decomposition for
viscoelasticity \cite{LatMonCM2015}.

The six-dimensional \emph{elastic}-corrector-type\ flow rule of Eq.
(\ref{flow rule logarithmic ct}) is to be compared to the nine-dimensional
\emph{plastic}-corrector-type\ flow rule given in Eq.
(\ref{flow rule mandel lp}) and its simplified version with $\mathbf{w}%
_{p}=\mathbf{0}$ of Eq. (\ref{flow rule mandelsym dp}). The
\emph{conventional} appearance of the elastic-corrector-type\ flow rule of Eq.
(\ref{flow rule logarithmic ct}) for \emph{anisotropic} elastoplasticity is
also to be compared to the \emph{non-conventional} appearance of the
elastic-corrector-type\ flow rule of Eq. (\ref{flow rule lie}) for
\emph{isotropic} elastoplasticity (which implicitly assumes $\mathbf{w}%
_{p}=\mathbf{0}$ as well). Clearly, Eq. (\ref{flow rule logarithmic ct})
yields the optimal computational parametrization (cf. Ref. \cite{LatMonCM2015}%
) for anisotropic multiplicative plasticity in the sense that will allow the
development of a new class of algorithms that exactly preserve the classical
return mapping schemes of the infinitesimal theory, hence circumventing
definitively the \textquotedblleft rate issue\textquotedblright\ \cite{Simo92}%
. In this respect, since $\Psi_{E}=\Psi_{E}\left(  \mathbf{E}_{e}\right)  $,
then Eq. (\ref{DissipationEe}) reads---note that the next interpretation is
possible due to the choice of $\mathbf{E}_{e}$ as the basic internal variable%
\begin{equation}
-\mathcal{D}=\mathbf{T}^{|e}:\,^{ct}\mathbf{\dot{E}}_{e}=\frac{d\Psi
_{E}\left(  \mathbf{E}_{e}\right)  }{d\mathbf{E}_{e}}:\,^{ct}\mathbf{\dot{E}%
}_{e}=\,^{ct}\dot{\Psi}_{E}<0\quad\text{if}\quad\dot{\gamma}>0
\label{-DissipationEe}%
\end{equation}
whereupon the dissipation rate is governed in the intermediate configuration
by the corrector logarithmic strain rate \emph{symmetric} tensor
$\,^{ct}\mathbf{\dot{E}}_{e}$ and its power conjugate generalized Kirchhoff
stress \emph{symmetric} tensor $\mathbf{T}^{|e}$, which follows the ideas
originally postulated by Eckart \cite{Eckart48}, Besseling \cite{Besseling66}
and Leonov \cite{Leonov76}, see Ref. \cite{Rubin16}. Remarkably, with the
present multiplicative formulation at hand, the thermodynamical stress tensor
that has traditionally governed the dissipation in the intermediate
configuration along with the \emph{non-symmetric} plastic deformation rate
tensor $\mathbf{l}_{p}$, i.e. the generally \emph{non-symmetric} Mandel stress
tensor $\mathbf{\Xi}^{|e}$ of Eq. (\ref{mandel S}) \cite{Mandel72},
\cite{Mandel74}, see Eq. (\ref{Dissipation mandel}), is not explicitly needed
any more.

\subsection{The \emph{stem} yield function}

We have seen that the dissipation equation, expressed in terms of correctors
elastic strain rates, may be written in any configuration and in terms of any
arbitrary pair of stress and strain work-conjugate measures. Their selections
are a matter of preference related to the stored energy function to be
employed and to the configuration where the yield function is to be defined.
It is not clear which one should be the stem configuration, i.e. the
configuration for which the tensor $\mathbb{N}$ is considered constant. We
coin herein this crucial aspect of the theory as the \textquotedblleft yield
function configuration issue\textquotedblright.

On one hand, it seems reasonable to choose the intermediate configuration as
the stem configuration so invariance is naturally obtained and $\mathbb{N}$
does not depend on the elastic strains or equivalently on the stress tensor.
On the other hand, using $\mathbb{N}_{S}$ as the tensor of constants in the
intermediate configuration results in a yield function in the current
configuration in terms of $\mathbf{\tau}^{|e}$ with nonorthogonal preferred
directions and depending of the elastic deformation through $\mathbb{N}_{\tau
}(\mathbf{X}_{e})$, cf. Eq. (\ref{potential taue Xe}).

Based on the understanding of the logarithmic strains evolution as the natural
generalization of the small strains one, see Ref. \cite{LatMonIJSS2014}, our
preference herein (as well as in Refs.
\cite{MontansBathe07,CamineroMontansBathe11}) are the internal elastic
logarithmic strains in the intermediate configuration $\mathbf{E}_{e}$ and
their work-conjugate internal generalized Kirchhoff stresses $\mathbf{T}^{|e}%
$, namely those governing the dissipation in Eqs. (\ref{DissipationEe}) and
(\ref{-DissipationEe}). Consistently, our preference is to choose
$\mathbb{N}_{T}$ as the specific tensor of yield constants associated to the
preferred material planes. Since $\mathbb{N}_{T}$ lies in the intermediate
configuration and $f_{T}(\mathbf{T}^{|e},k)$ is written in terms of (material)
generalized elastic Kirchhoff stresses, its natural push-forward to the
current configuration (performed with the elastic rotations $\mathbf{R}_{e}$)
leads to a yield function in terms of the (spatial) generalized elastic
Kirchhoff stresses that preserves the orthogonality of the main material
directions in $\mathbb{N}_{T}$ and that is still constant in the elastically
rotated frame. We further note that when loading in principal material axes or
considering elastic isotropy (even with plastic anisotropy) the generalized
elastic Kirchhoff stresses $\mathbf{T}^{|e}$ are the rotated elastic Kirchhoff
stresses $\mathbf{\tau}^{|e}$ of Eq. (\ref{rotated kirchhoff stresses})
\cite{LatMonAPM2016}. Furthermore, the numerical integration of the flow rule
of Eq. (\ref{flow rule logarithmic ct}) may be directly performed with a
backward-Euler additive scheme, without explicitly employing exponential
mappings, and plastic volume preservation is automatically accomplished for
models of plasticity possessing a pressure insensitive yield criterion, hence
rendering the most natural generalization of the classical return mapping
algorithms of the infinitesimal theory \cite{LatMonCM2015}.

Proceeding exactly as in both the small strain case and the finite strain
spatial framework, we identify in Eq. (\ref{Dissipation Te}) the following
yield function $f_{T}(\mathbf{T}^{|e},k)$ and the loading/unloading
conditions, i.e.%
\begin{equation}
f_{T}(\mathbf{T}^{|e},k)=2\phi_{T}(\mathbf{T}^{|e})-k^{2}=\mathbf{T}%
^{|e}:\mathbb{N}_{T}:\mathbf{T}^{|e}-k^{2}=0\quad\text{if}\quad\dot{\gamma}>0
\label{Yield fcn log}%
\end{equation}
and%
\begin{equation}
\dot{\gamma}=0\quad\text{if}\quad f_{T}(\mathbf{T}^{|e},k)=2\phi
_{T}(\mathbf{T}^{|e})-k^{2}=\mathbf{T}^{|e}:\mathbb{N}_{T}:\mathbf{T}%
^{|e}-k^{2}<0
\end{equation}
whereupon we obtain the dissipation in terms of the (characteristic) internal
flow stress $k>0$ and the (characteristic) frictional deformation rate
$\dot{\gamma}\geq0$ as%
\begin{equation}
\mathcal{D}=k\dot{\gamma}\geq0\text{\quad for\quad}\dot{\gamma}\geq0
\end{equation}

\subsubsection{Change of stress measures and configuration}

The yield function may be written also in the reference or current
configurations or as a function of any other stress measure, still being
exactly the same yield condition. For example, the potential $\phi
_{T}(\mathbf{T}^{|e})$ may be expressed in terms of the second
Piola--Kirchhoff stresses $\mathbf{S}^{|e}$ of Eq. (\ref{S|e}) using---the
fourth-order tensor $\mathbb{M}_{\dot{E}_{e}}^{\dot{A}_{e}}$ maps both
$\mathbf{\dot{E}}_{e}$ to $\mathbf{\dot{A}}_{e}$ and, by power invariance,
$\mathbf{S}^{|e}$ to $\mathbf{T}^{|e}$ \cite{LatMonAPM2016}%
\begin{equation}
\mathbf{T}^{|e}=\mathbf{S}^{|e}:\frac{d\mathbf{A}_{e}}{d\mathbf{E}_{e}%
}=\mathbf{S}^{|e}:\mathbb{M}_{\dot{E}_{e}}^{\dot{A}_{e}}%
\end{equation}
so---we note that $\mathbb{M}_{\dot{E}_{e}}^{\dot{A}_{e}}$ has major
symmetries and only depends on the spectral decomposition of the elastic right
stretch tensor $\mathbf{U}_{e}$ \cite{LatMonAPM2016} and that $\mathbf{U}_{e}$
does not represent a change of the reference configuration since
$\mathbf{T}^{|e}$ and $\mathbf{S}^{|e}$ lie in the same placement%
\begin{equation}
\phi_{T}(\mathbf{T}^{|e})=\tfrac{1}{2}\mathbf{T}^{|e}:\mathbb{N}%
_{T}:\mathbf{T}^{|e}=\tfrac{1}{2}\mathbf{S}^{|e}:\mathbb{N}_{S}\left(
\mathbf{U}_{e}\right)  :\mathbf{S}^{|e}=\phi_{S}(\mathbf{S}^{|e}%
,\mathbf{U}_{e})
\end{equation}
with%
\begin{equation}
\mathbb{N}_{S}\left(  \mathbf{U}_{e}\right)  :=\mathbb{M}_{\dot{E}_{e}}%
^{\dot{A}_{e}}:\mathbb{N}_{T}:\mathbb{M}_{\dot{E}_{e}}^{\dot{A}_{e}}%
\end{equation}
In the spatial configuration, we can similarly write%
\begin{equation}
f_{\tau}(\mathbf{\tau}^{|e},k;\mathbf{X}_{e})=\mathbf{\tau}^{|e}%
:\mathbb{N}_{\tau}\left(  \mathbf{X}_{e}\right)  :\mathbf{\tau}^{|e}%
-k^{2}=0\quad\text{if}\quad\dot{\gamma}>0
\end{equation}
with---the fourth-order tensor $\mathbb{M}_{\dot{E}_{e}}^{d_{e}}$ maps both
$\mathbf{\dot{E}}_{e}$ to $\mathbf{d}_{e}$ and, by power invariance,
$\mathbf{\tau}^{|e}$ to $\mathbf{T}^{|e}$ \cite{LatMonAPM2016}%
\begin{equation}
\mathbb{N}_{\tau}\left(  \mathbf{X}_{e}\right)  =\mathbb{M}_{\dot{E}_{e}%
}^{d_{e}}(\mathbf{X}_{e}):\mathbb{N}_{T}:\mathbb{M}_{\dot{E}_{e}}^{d_{e}%
}(\mathbf{X}_{e}) \label{Ntau NT}%
\end{equation}
However, if, for example, $\mathbb{N}_{T}$ is a fourth-order tensor of yield
constants when is represented in the preferred material directions in the
intermediate configuration, then $\mathbb{N}_{S}\left(  \mathbf{U}_{e}\right)
=\mathbb{M}_{\dot{E}_{e}}^{\dot{A}_{e}}:\mathbb{N}_{T}:\mathbb{M}_{\dot{E}%
_{e}}^{\dot{A}_{e}}$ will change with the \emph{elastic} strains (which for
the case of metals are assumed to be small and could be arguably neglected for
this purpose), and vice-versa. Note also that once the stem configuration has
been decided, $k$ is the same constant for any case and that the dissipation
$\mathcal{D}=k\dot{\gamma}$ is of course an invariant value independent also
of the chosen stress/strain couple.

\subsubsection{Other possible yield functions}

The form of the yield function of Eq. (\ref{Yield fcn log}) includes just some
of the possibilities. Other more general possibilities may be considered. For
example, assume the potential%
\begin{equation}
\phi_{T}=\tfrac{1}{2}\mathbf{T}^{|e}:\mathbb{N}_{T}:\mathbf{T}^{|e}%
+\mathbf{N}_{T}:\mathbf{T}^{|e}%
\end{equation}
where $\mathbf{N}_{T}$ is a second order tensor. Then Eq.
(\ref{flow rule logarithmic ct}) yields%
\begin{equation}
\,^{ct}\mathbf{\dot{E}}_{e}=-\dot{\gamma}\frac{1}{k}(\mathbb{N}_{T}%
:\mathbf{T}^{|e}+\mathbf{N}_{T})
\end{equation}
and Eq. (\ref{DissipationEe}) gives%
\begin{equation}
\mathcal{D}=\dot{\gamma}\frac{1}{k}(\mathbf{T}^{|e}:\mathbb{N}_{T}%
:\mathbf{T}^{|e}+\mathbf{N}_{T}:\mathbf{T}^{|e})>0\quad\text{if}\quad
\dot{\gamma}>0
\end{equation}
where we identify the yield function%
\begin{equation}
\bar{f}_{T}:=\mathbf{T}^{|e}:\mathbb{N}_{T}:\mathbf{T}^{|e}+\mathbf{N}%
_{T}:\mathbf{T}^{|e}-k^{2}=0\quad\text{if}\quad\dot{\gamma}>0
\end{equation}
so%
\begin{equation}
\mathcal{D}=k\dot{\gamma}\geq0\text{\quad for\quad}\dot{\gamma}\geq0
\end{equation}

For example if $\mathbb{N}_{T}=\mathbb{P}^{S}$ is the fourth-order deviatoric
projection tensor in the logarithmic strain space (i.e. the same one as in the
small strain case) and $\mathbf{N}_{T}=\mathbf{0}$, then we recover a
von-Mises-like yield surface defined in terms of the stresses $\mathbf{T}%
^{|e}$ in the intermediate configuration. For the case of $\mathbf{N}%
_{T}=\mathbf{0}$ and $\mathbb{N}_{T}$ a fourth-order orthotropic deviatoric
tensor, then we obtain a Hill-like yield criterion. For the case
$\mathbb{N}_{T}=\mathbb{P}^{S}$ and $\mathbf{N}_{T}=\alpha\mathbf{I}$, with
$\alpha$ being a scalar, we obtain a Drucker-Prager-like yield criterion
\cite{KojicBathe87,Borjabook}. And so forth. Of course, non-associative flow
rules are possible as well (cf. the equivalent Eqs.
(\ref{flow rule nonassociative}) and (\ref{dissipation nonassociative})), but
then positive dissipation and symmetric response linearization are not
guaranteed, as it is known \cite{Simo98}.

\subsection{Determination of model \emph{internal} parameters}

The internal stress tensor $\mathbf{T}^{|e}$, as given in Eq. (\ref{Tinte}),
is defined in the intermediate configuration, hence it is not measurable. This
means that the specific form of the constitutive relations, especially of the
yield condition, is built up with non-measurable quantities. We show in this
section that the internal parameters of the selected model can be obtained
from experimental testing in any case. We address the yield function
determination as an example.

Consider the internal yield function given in Eq. (\ref{Yield fcn log}). The
corresponding external stress tensor is given by Eq. (\ref{TTextTint}). Assume
now that we want to determine the Hill-type yield function parameters,
included in the fourth order tensor $\mathbb{N}_{T}$, and the internal flow
stress $k$ from experimental tests. We consider a uniaxial test performed over
a preferred axis of the corresponding orthotropic material at hand. Since
there are no rotations present, all the strain tensors (elastic, plastic and
total) are coaxial so logarithmic strains are additive, i.e.%
\begin{equation}
\mathbf{X}=\mathbf{U}=\mathbf{U}_{e}\mathbf{U}_{p}\quad\Rightarrow
\quad\mathbf{E}=\mathbf{E}_{e}+\mathbf{E}_{p}%
\end{equation}
so the general relation $\mathbf{E}_{e}\left(  \mathbf{E},\mathbf{X}%
_{p}\right)  $ specifies for this particular case to%
\begin{equation}
\mathbf{E}_{e}=\mathbf{E}-\mathbf{E}_{p}%
\end{equation}
The purely kinematical mapping tensor present in Eq. (\ref{TTextTint})
particularizes to the fourth-order identity tensor%
\begin{equation}
\left.  \frac{\partial\mathbf{E}_{e}}{\partial\mathbf{E}}\right\vert _{%
\xpdo
}=\left.  \frac{\partial\left(  \mathbf{E}-\mathbf{E}_{p}\right)  }%
{\partial\mathbf{E}}\right\vert _{%
\xpdo
}=\frac{\partial\mathbf{E}}{\partial\mathbf{E}}=\mathbb{I}%
\end{equation}
and the external stress $\mathbf{T}$ during the uniaxial test reduces to%
\begin{equation}
\mathbf{T}=\mathbf{T}^{|e}%
\end{equation}
Therefore, the yield function during the uniaxial test is exactly recast as%
\begin{equation}
f(\mathbf{T}^{|e},k)\equiv f(\mathbf{T},k)=\mathbf{T}:\mathbb{N}%
_{T}:\mathbf{T}-k^{2}%
\end{equation}
Furthermore, the generalized Kirchhoff stress tensor $\mathbf{T}$, which is
work-conjugate of the logarithmic strain tensor, is coincident with the
Kirchhoff stress tensor $\mathbf{\tau}$ for rotationless cases along preferred
directions \cite{LatMonAPM2016}. Thus we also have the identity%
\begin{equation}
f(\mathbf{T}^{|e},k)\equiv f(\mathbf{T},k)\equiv f(\mathbf{\tau}%
,k)=\mathbf{\tau}:\mathbb{N}_{T}:\mathbf{\tau}-k^{2} \label{Yield fcn tau}%
\end{equation}
and the yield function becomes expressed in terms of stress quantities being
fully measurable. When yielding takes place%
\begin{equation}
\mathbf{\tau}=\mathbf{\tau}_{y}%
\end{equation}
is known, where $\mathbf{\tau}_{y}$ includes the corresponding Kirchhoff flow
stress components, and also $f(\mathbf{\tau},k)=0$.

It can be shown that similar expressions hold for shear tests within material
preferred planes, where the purely kinematical internal-to-external mapping,
relating internal stresses to external stresses, is always known at each
deformation state. Hence, the fourth order tensor $\mathbb{N}_{T}$ and the
internal yield function parameter $k$, that define the internal yield
function, can be completely determined from the proper number of measured
experimental data.

Finally, this yield function can be used in further calculations involving
general three-dimensional deformation states, because in these cases we always
know the internal strain $\mathbf{E}_{e}$ obtained from the Lee decomposition,
and consequently $\mathbf{T}^{|e}$.

\begin{table}[h]
\caption{Finite strain multiplicative anisotropic elastoplasticity model
formulated in terms of logarithmic strains in the intermediate configuration.}%
\label{Table - Logarithmic}
\begin{center}%
\begin{tabular}
[c]{|l|}\hline%
\begin{tabular}
[c]{l}%
\vspace{-0.1in}\\
(i) Multiplicative decomposition of the deformation gradient $\mathbf{X}%
=\mathbf{X}_{e}\mathbf{X}_{p}$\bigskip\\
(ii) Symmetric internal strain variable $\mathbf{E}_{e}=\frac{1}{2}%
\ln(\mathbf{X}_{e}^{T}\mathbf{X}_{e})$\bigskip\\
(iii) Kinematics induced by $\mathbf{E}_{e}(\mathbf{E},\mathbf{X}_{p}%
)$\medskip\\
\qquad$\mathbf{\dot{E}}_{e}=\left.  \mathbf{\dot{E}}_{e}\right\vert
_{{\scriptsize \mathbf{\dot{X}}_{p}=\mathbf{0}}}+\left.  \mathbf{\dot{E}}%
_{e}\right\vert _{{\scriptsize \mathbf{\dot{E}}=\mathbf{0}}}=\,^{tr}%
\mathbf{\dot{E}}_{e}+\,^{ct}\mathbf{\dot{E}}_{e}\neq\mathbf{\dot{E}%
}-\mathbf{\dot{E}}_{p}$\bigskip\\
(iv) Symmetric stresses deriving from the strain energy $\Psi_{E}%
(\mathbf{E}_{e})$\medskip\\
\qquad$\mathbf{T}^{|e}=\dfrac{d\Psi_{E}(\mathbf{E}_{e})}{d\mathbf{E}_{e}}%
$\,,\qquad$\mathbf{T}=\dfrac{\partial\Psi_{E}(\mathbf{E},\mathbf{X}_{p}%
)}{\partial\mathbf{E}}=\mathbf{T}^{|e}:\dfrac{\partial\mathbf{E}%
_{e}(\mathbf{E},\mathbf{X}_{p})}{\partial\mathbf{E}}\neq\mathbf{T}^{|e}%
$\bigskip\\
(v) Evolution equation for associative symmetric plastic flow\medskip\\
\qquad$-\,^{ct}\mathbf{\dot{E}}_{e}=\dot{\gamma}\dfrac{1}{k}\nabla\phi
_{T}(\mathbf{T}^{|e})\neq\mathbf{\dot{E}}_{p}$\medskip\\
\qquad$\dot{\gamma}\geq0$\,,\quad$f_{T}(\mathbf{T}^{|e},k)=2\phi
_{T}(\mathbf{T}^{|e})-k^{2}\leq0$\,,\quad$\dot{\gamma}f_{T}(\mathbf{T}%
^{|e},k)=0$\bigskip\\
(vi) Additional evolution equation for skew-symmetric plastic flow
$\mathbf{w}_{p}$\medskip\\
{\small {Note: Potential $\Psi_{E}(\mathbf{E}_{e})$ and function
$f_{T}(\mathbf{T}^{|e},k)$ are anisotropic, in general.}\bigskip}\\
\end{tabular}
\\\hline
\end{tabular}
\end{center}
\end{table}

\begin{table}[h]
\caption{Finite strain multiplicative anisotropic elastoplasticity model
formulated in terms of quadratic strains in the intermediate configuration.}%
\label{Table - Quadratic}
\begin{center}%
\begin{tabular}
[c]{|l|}\hline%
\begin{tabular}
[c]{l}%
\vspace{-0.1in}\\
(i) Multiplicative decomposition of the deformation gradient $\mathbf{X}%
=\mathbf{X}_{e}\mathbf{X}_{p}$\bigskip\\
(ii) Symmetric internal strain variable $\mathbf{A}_{e}=\frac{1}{2}%
(\mathbf{X}_{e}^{T}\mathbf{X}_{e}-\mathbf{I})$\bigskip\\
(iii) Kinematics induced by $\mathbf{A}_{e}(\mathbf{A},\mathbf{X}%
_{p})=\mathbf{X}_{p}^{-T}(\mathbf{A}-\mathbf{A}_{p})\mathbf{X}_{p}^{-1}%
$\medskip\\
\qquad$\mathbf{\dot{A}}_{e}=\left.  \mathbf{\dot{A}}_{e}\right\vert
_{{\scriptsize \mathbf{\dot{X}}_{p}=\mathbf{0}}}+\left.  \mathbf{\dot{A}}%
_{e}\right\vert _{{\scriptsize \mathbf{\dot{E}}=\mathbf{0}}}=\,^{tr}%
\mathbf{\dot{A}}_{e}+\,^{ct}\mathbf{\dot{A}}_{e}\neq\mathbf{\dot{A}%
}-\mathbf{\dot{A}}_{p}$\bigskip\\
(iv) Symmetric stresses deriving from the strain energy $\Psi_{A}%
(\mathbf{A}_{e})$\medskip\\
\qquad$\mathbf{S}^{|e}=\dfrac{d\Psi_{A}(\mathbf{A}_{e})}{d\mathbf{A}_{e}}%
$\,,\quad$\mathbf{S}=\dfrac{\partial\Psi_{A}(\mathbf{A},\mathbf{X}_{p}%
)}{\partial\mathbf{A}}=\mathbf{S}^{|e}:\dfrac{\partial\mathbf{A}%
_{e}(\mathbf{A},\mathbf{X}_{p})}{\partial\mathbf{A}}=\mathbf{X}_{p}%
^{-1}\mathbf{S}^{|e}\mathbf{X}_{p}^{-T}$\bigskip\\
(v) Evolution equation for associative symmetric plastic flow\medskip\\
\qquad$-\,^{ct}\mathbf{\dot{A}}_{e}=\dot{\gamma}\dfrac{1}{k}\nabla\phi
_{S}(\mathbf{S}^{|e})\neq\mathbf{\dot{A}}_{p}$\medskip\\
\qquad$\dot{\gamma}\geq0$\,,\quad$f_{S}(\mathbf{S}^{|e},k)=2\phi
_{S}(\mathbf{S}^{|e})-k^{2}\leq0$\,,\quad$\dot{\gamma}f_{S}(\mathbf{S}%
^{|e},k)=0$\bigskip\\
(vi) Additional evolution equation for skew-symmetric plastic flow
$\mathbf{w}_{p}$\medskip\\
{\small {Note: Potential $\Psi_{A}(\mathbf{A}_{e})$ and function
$f_{S}(\mathbf{S}^{|e},k)$ are anisotropic, in general.}\bigskip}\\
\end{tabular}
\\\hline
\end{tabular}
\end{center}
\end{table}

\section{Numerical example}

In this example we simulate numerically three cyclic tension-compression
uniaxial tests along orthotropy material axes in order to show that the
logarithmic-based model reproduces some basic elastoplastic responses within
an incompressible orthotropic finite strain context. The integration of the
corrector-elastic-type flow rule of Eq. (\ref{flow rule logarithmic ct}) is
performed during plastic steps employing a simple backward-Euler additive
formula, see details in Ref. \cite{LatMonCM2015}. In this elastoplasticity
case, the yield condition fulfillment is an additional constraint to be
imposed during local iterations.

We consider an additive uncoupled decomposition for the total strain energy
function $\Psi_{E}\left(  \mathbf{E}_{e}\right)  =\mathcal{W}(\mathbf{E}%
_{e}^{d})+\mathcal{U}(\mathbf{E}_{e}^{v})$ in terms of its purely deviatoric
and volumetric parts, respectively, where $\mathbf{E}_{e}^{v}=\frac{1}%
{3}tr(\mathbf{E}_{e})\mathbf{I}=\frac{1}{3}\ln(J_{e})\mathbf{I}$ is the
volumetric elastic strain tensor, with $J_{e}=\det\mathbf{X}_{e}$ the elastic
Jacobian and $\mathbf{I}$ the second-order identity tensor, and $\mathbf{E}%
_{e}^{d}=\mathbf{E}_{e}-\mathbf{E}_{e}^{v}$ is the distortional one, cf. for
example Ref. \cite{LatMonCM2014}. We define the following deviatoric strain
energy function---the volumetric \emph{penalty} function is taken stiff enough
so that elastic incompressibility ($J_{e}\rightarrow1$) is numerically imposed
during the computations%
\begin{equation}
\mathcal{W}(\mathbf{E}_{e}^{d})=\mu_{1}(E_{e1}^{d})^{2}+\mu_{2}(E_{e2}%
^{d})^{2}+\mu_{3}(E_{e3}^{d})^{2}=5(E_{e1}^{d})^{2}+3(E_{e2}^{d})^{2}%
+2(E_{e3}^{d})^{2}~%
\operatorname{MPa}%
\label{W}%
\end{equation}
where only its axial components in preferred material directions are needed
for this specific example. In order to complete the definition of the model
within preferred axes $X_{pr}$, we assume a Hill-type pressure-insensitive
yield function with no hardening. The yield function of Eq.
(\ref{Yield fcn log}) simplifies to Eq. (\ref{Yield fcn tau}) with
$\mathbb{N}_{T}=\mathbb{P}^{S}:\mathbb{\bar{N}}_{T}:\mathbb{P}^{S}$, where
$\mathbb{\bar{N}}_{T}$ is a fourth-order \textquotedblleft
diagonal\textquotedblright\ tensor (when it is represented in matrix, Voigt
notation in preferred directions) containing independent yielding weight
factors \cite{KojicBatheBook} and $\mathbb{P}^{S}$ is the fourth-order
deviatoric projection tensor. Only the axial-to-axial components of the matrix
representation of the tensor $\mathbb{N}_{T}$ are needed for in-axes loading
cases, so we consider the left-upper $3\times3$ matrix blocks of the
respective $6\times6$ symmetric matrices. We just take for this representative
example%
\begin{equation}
\left[  \mathbb{N}_{T}\right]  _{X_{pr}}=\left[
\begin{array}
[c]{ccc}%
\frac{2}{3} & -\frac{1}{3} & -\frac{1}{3}\\
-\frac{1}{3} & \frac{2}{3} & -\frac{1}{3}\\
-\frac{1}{3} & -\frac{1}{3} & \frac{2}{3}%
\end{array}
\right]  \left[
\begin{array}
[c]{ccc}%
1 & 0 & 0\\
0 & 2 & 0\\
0 & 0 & 3
\end{array}
\right]  \left[
\begin{array}
[c]{ccc}%
\frac{2}{3} & -\frac{1}{3} & -\frac{1}{3}\\
-\frac{1}{3} & \frac{2}{3} & -\frac{1}{3}\\
-\frac{1}{3} & -\frac{1}{3} & \frac{2}{3}%
\end{array}
\right]  \label{Nt}%
\end{equation}
and also prescribe $k=k_{0}=10%
\operatorname{MPa}%
$ in Eq. (\ref{Yield fcn tau}).

From the strain energy of Eq. (\ref{W}) we can analytically calculate the
preferred Young moduli \cite{LatMonCM2015}%
\begin{equation}
Y_{1}=62/5=12.4%
\operatorname{MPa}%
~\text{,\quad}Y_{2}=62/7=8.857%
\operatorname{MPa}%
~\text{,\quad}Y_{3}=31/4=7.75%
\operatorname{MPa}
\label{Young}%
\end{equation}
On the other side, Equation (\ref{Yield fcn tau}) with $k=k_{0}=10%
\operatorname{MPa}%
$ and the axial-to-axial components of $\mathbb{N}_{T}$ given in Eq.
(\ref{Nt}), specialized for the three tests separately gives the following
yield stresses as result---note additionally that Cauchy stresses are
coincident with Kirchhoff stresses by incompressibility%
\begin{equation}
\sigma_{y1}=10%
\operatorname{MPa}%
~\text{,\quad}\sigma_{y2}=5\sqrt{3}=8.66%
\operatorname{MPa}%
~\text{,\quad}\sigma_{y3}=2\sqrt{15}=7.746%
\operatorname{MPa}
\label{yieldStresses}%
\end{equation}

\begin{figure}[h]
\begin{center}
\includegraphics[width=0.55\textwidth]{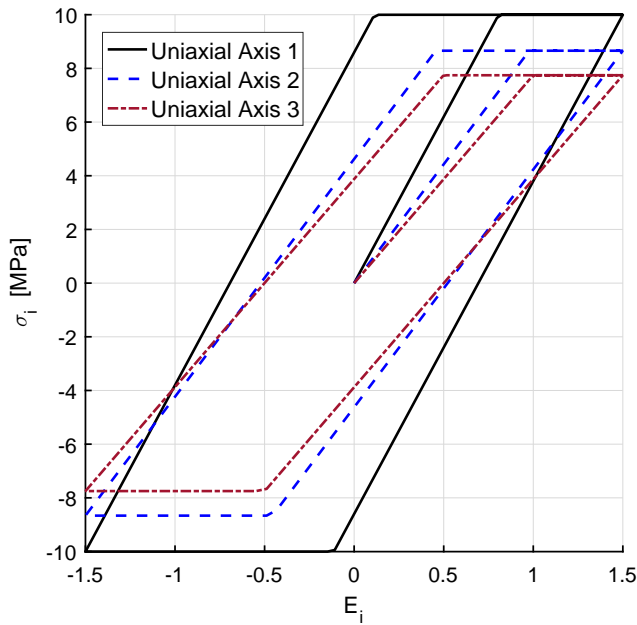}
\end{center}
\caption{Cyclic tension-compression uniaxial tests over orthotropy preferred
directions. We represent by $\sigma_{i}$ and $E_{i}$ the uniaxial components
of the Cauchy stress and the logarithmic strain in the test performed in axis
$(i)$. Perfect plasticity case, i.e. $k=k_{0}=const$.}%
\label{Figure - Figure_Example.eps}%
\end{figure}

We can verify in Figure \ref{Figure - Figure_Example.eps} that the values of
Eqs. (\ref{Young}) and (\ref{yieldStresses}), which have been calculated
analytically, are effectively reproduced by the simulations, for which only
the internal model parameters $\mu_{1}$, $\mu_{2}$, $\mu_{3}$, $k$,
$(\mathbb{\bar{N}}_{T})_{22}/(\mathbb{\bar{N}}_{T})_{11}=2$ and $(\mathbb{\bar
{N}}_{T})_{33}/(\mathbb{\bar{N}}_{T})_{11}=3$ have been defined. We can also
observe that a perfect plasticity case, i.e. with no hardening, is obtained
and that both elastic and plastic strains are large.

\section{Conclusion}

In this paper we have presented a novel framework for elastoplasticity at
large strains. This framework, grounded in the multiplicative decomposition,
naturally solves the \textquotedblleft rate issue\textquotedblright; i.e. the
flow rule is naturally obtained in terms of a corrector elastic strain rate
which simply results to be a partial contribution to the total rate of such
strain, exactly as in the small strain theory. The new approach results in
essentially the same type of equations in small strains and in large strains,
and whether the latter are integrated in the intermediate or in the spatial
configurations. The \emph{continuum }framework also naturally results in the
typical two stages of the \emph{algorithmic} integration of elastoplastic
equations: the trial elastic predictor and the plastic corrector. Hence the
development of integration algorithms employing this proposal is
straightforward by the direct use of the backward-Euler integration rule over
the corrector logarithmic strain rate without explicitly employing exponential
mappings. The large strain formulation, being simpler than most proposals in
the literature, is also general, meaning that it is not restricted to moderate
elastic strains and it is not restricted to isotropy. Furthermore, as shown in
the manuscript, there is no need to perform any dissipation hypothesis in the
plastic spin, which remains uncoupled and completely independent of the
integration of the symmetric flow. The present formulation may be equally
employed in metal plasticity or in the plastic behavior of soft materials.

\section*{Acknowledgements}

Partial financial support for this work has been given by grants DPI2011-26635
and DPI2015-69801-R from the Direcci\'{o}n General de Proyectos de
Investigaci\'{o}n of the Ministerio de Econom\'{\i}a y Competitividad of
Spain. F.J. Mont\'{a}ns also acknowledges the support of the Department of
Mechanical and Aerospace Engineering of University of Florida during the
sabbatical period in which this paper was finished and that of the Ministerio
de Educaci\'{o}n, Cultura y Deporte of Spain for the financial support for
that stay under grant PRX15/00065.

\end{document}